\documentclass[a4paper,11pt]
{article}
\usepackage{jheppub}

\usepackage{amsmath,amssymb}
\usepackage{graphicx}
\usepackage{listings}
\usepackage{xcolor}
\hypersetup{colorlinks=true,citecolor=green,linkcolor=green,urlcolor=green}

\title{\textbf{Flavor-Modulated Neutrino Communication: Channel Model, Modulation Schemes, and Link Budget for Through-Earth Links}}

\author{Bipin Singh Koranga}
\emailAdd{bskoranga@kmc.du.ac.in}
\affiliation{Department of Physics, Kirori Mal College, University of Delhi, Delhi 110007, India}
\author{Biswmkwr Boro}
\emailAdd{borobiswmkwr2@gmail.com}
\affiliation{Department of Physics, Kirori Mal College, University of Delhi, Delhi 110007, India}
\author{Imran Khan}
\affiliation{Department of Physics, Jamia Millia Islamia, New Delhi 110025, India}

\author{Vivek Kumar Nautiyal}
\emailAdd{viveknautiyal01@gmail.com}
\affiliation{Department of Physics, Chaudhary Charan Singh University, Meerut - 250004, India}

\abstract{We investigate the physical limits of transmitting information through the Earth using neutrinos, modeling the link as a discrete-input, Poisson-output classical channel whose crosstalk is set by the three-flavor evolution operator in a layered, matter-filled Earth. For the maximum-likelihood receiver, we derive closed-form Bhattacharyya and Chernoff error exponents for Poisson counts, accurate to within 0.4\% of the exact Chernoff limit. We examine three modulation schemes: source-flavor keying (SFK), oscillation-phase keying (OPK), and matter-path keying (MPK). For an Earth-diameter link with a neutrino-factory-class source (\(3 \times 10^{13}\) decays/s) and a 200 kt detector, the mean charged-current event rate is \(1.31 \times 10^{-2}\,\mathrm{s^{-1}}\), yet the information rate is only \(1.36 \times 10^{-4}\) bit/s for SFK and \(5.2 \times 10^{-6}\) bit/s for OPK. This reduction occurs because at \(E \simeq 25\) GeV the first oscillation maximum nearly fully relabels \(\nu_\mu \to \nu_\tau\) (\(T_{\mu\tau} \simeq 0.98\)); SFK relies mainly on a count-based \(e\)-like channel, while OPK suffers from suppressed appearance probabilities. For DUNE and T2HK with horn-polarity keying, assuming \(10^{4}\) unoscillated events per year, rates are \(\sim 10^{-6}\)--\(10^{-7}\) bit/s. The OPK estimate is highly sensitive to the Earth density profile: full tabulated PREM lowers it threefold relative to a three-layer model, and a \(\pm 2\%\) density rescaling changes it by an order of magnitude, whereas SFK varies by less than 10\%. Flavor modulation does not remove the fundamental flux limit; it adds a signaling dimension with constant-intensity operation and directional access, though these do not constitute a cryptographic guarantee.}

\keywords{Neutrino Physics, Phenomenological Models, Information Theory}

\begin{document}

\maketitle

\section{Introduction}\label{s1}

Communication through dense media is limited by absorption. Radio waves are attenuated within metres of seawater or rock, and seismic and acoustic links are restricted in range and bandwidth. The weakly interacting neutrino is a natural candidate carrier for such links, because a GeV-scale neutrino traverses the Earth diameter, \(D_{\oplus} = 12742\) km, with an optical depth of order \(10^{- 3}\) (Sec.~\ref{s2.4}). The price is that neutrinos are also hard to produce in quantity and hard to detect: the carrier needs a facility-scale source and a kiloton-scale receiver.

A proof of principle was given in 2012, when a short text message was sent through about a kilometre of rock using the NuMI beam and the MINERvA detector, with a rate of order 0.1 bit/s and a low error rate \cite{ref1}. That demonstration used the simplest possible modulation, on--off keying (OOK): a neutrino pulse means ``1'' and its absence means ``0''. The information then resides only in the intensity of the beam. Neutrinos, however, carry internal degrees of freedom that intensity does not use. Since the discovery of oscillations \cite{ref2,ref3}, the flavor state of a neutrino beam is known to evolve with the ratio \(L/E\) of baseline to energy and, in matter, with the electron density through the Mikheyev--Smirnov--Wolfenstein (MSW) effect \cite{ref6,ref7}. The flavor composition at the source, the beam energy, the particle type (neutrino or antineutrino) and the path through the Earth are all controllable by the transmitter and measurable by the receiver. The question we address is how much information these extra dimensions can carry, and at what cost.

We emphasize the scope of this work. It is a phenomenological study at the interface of oscillation physics and communication theory, and it is not a proposal for a practical near-term telecommunication system. On the oscillation side, the mixing angles and mass-squared splittings are determined by global fits, and the transition matrix \(T_{\alpha\beta}(E,P)\) along a given path is a calculable object with quantified parametric uncertainty. Communication theory supplies the language of channels, capacities and error exponents in which the resulting limits can be stated precisely. The rates we obtain, \(\sim 10^{-4}\) bit/s for an Earth-diameter link, make plain that the value of the subject lies not in throughput, but in the physical structure of the channel. Our treatment is strictly classical: oscillation is a coherent quantum process, but the detection of a neutrino flavor is a classical measurement outcome. We make no claim of quantum-capacity advantage, entanglement-assisted signaling or superluminal communication.

The accuracy of any through-Earth link budget rests on the density profile along the chord. The dependence of long-baseline oscillation probabilities on Earth-density modelling has been studied previously \cite{ref16}, and anomalous matter potentials in the Earth's inner core, as a probe of dark matter and light sterile neutrinos, in Ref.~\cite{ref17}; both bear directly on the appearance-based schemes analyzed here.

The statistical structure of the problem is that of a photon-counting channel. Detection of neutrino events in a symbol interval is a Poisson process whose mean is fixed by the source, the oscillation probabilities and the detector response. What has been missing is a channel model that connects the oscillation operator in a realistic, density-varying Earth to these information-theoretic tools, so that modulation schemes can be compared by error exponents, not by qualitative argument. 

In this paper, we build this channel from first principles: the three-flavor evolution operator in a layered Earth defines a doubly stochastic flavor transition matrix, and the detector confusion matrix converts it into a Poisson-output channel (Secs.~\ref{s3} and~\ref{s4}). We derive the maximum-likelihood receiver, closed-form Bhattacharyya and Chernoff error exponents, and formulate flavor modulation in detail across source flavor, energy, particle type, and path (Secs.~\ref{s5} and~\ref{s6}). Finally, we evaluate the link budget numerically and test its sensitivity to the Earth density profile by comparing a three-layer model with the full tabulated PREM (Secs.~\ref{s7} and~\ref{s8}).

Our main quantitative findings are as follows. For an Earth-diameter link with a neutrino-factory-class source of \(3 \times 10^{13}\) decays per second and a 200 kt detector, the mean charged-current rate is \(1.31 \times 10^{-2}\,\mathrm{s^{-1}}\). At \(E \simeq 25\) GeV, this corresponds to the first oscillation maximum where matter suppresses \(\nu_{\mu} \rightarrow \nu_{e}\) conversion and drives a near-total relabeling \(\nu_\mu \to \nu_\tau\) (\(T_{\mu\tau} \simeq 0.98\)). Consequently, the information rate is only \(1.36 \times 10^{-4}\) bit/s for source-flavor keying (SFK) and \(5.2 \times 10^{-6}\) bit/s for oscillation-phase keying (OPK). Applying this framework to planned long-baseline facilities (DUNE, T2HK) with horn-polarity keying yields \(\sim 10^{-6}\)--\(10^{-7}\) bit/s for an assumed yield of \(10^{4}\) unoscillated events per year. 

Crucially, we find that the OPK estimate is highly sensitive to the Earth density profile; a full tabulated-PREM evaluation lowers it by a factor of about three relative to a three-layer model, and a \(\pm 2\%\) rescaling of the density changes it by an order of magnitude. The SFK estimate remains robust under these variations. Flavor modulation therefore does not remove the flux limit, which is set by the product of source intensity and detector mass. It adds a second signaling dimension, permits constant-intensity operation, and makes the link directional. 

The paper is organized as follows. Section~\ref{s2} summarizes oscillations, matter effects, Earth chords and transparency. Section~\ref{s3} defines the channel and the receiver, and Sec.~\ref{s4} gives the information-theoretic bounds. Section~\ref{s5} develops flavor modulation in detail, and Sec.~\ref{s6} specifies the modulation schemes. Sections~\ref{s7} and \ref{s8} give the link budget for an Earth-diameter link and for planned facilities. Practical limitations are discussed in Sec.~\ref{s9}, and Sec.~\ref{s10} concludes.

\section{Physical Background}\label{s2}

\subsection{Vacuum oscillations}\label{s2.1}

Flavor and mass states are related by \(|\nu_\alpha\rangle=\sum_i U^*_{\alpha i}|\nu_i\rangle\), with \(U\) the PMNS matrix \cite{ref4,ref5}. In the flavor basis the Hamiltonian is

\begin{equation}
\label{eq:1}
H_{\rm vac}=\frac{1}{2E}\,U\,\mathrm{diag}\!\left(0,\Delta m^2_{21},\Delta m^2_{31}\right)U^\dagger,
\end{equation}

and the transition probability is

\begin{equation}
\label{eq:2}
\begin{split}
P_{\alpha\beta}&=\Bigl|\sum_i U^*_{\alpha i}U_{\beta i}\,e^{-i m_i^2 L/2E}\Bigr|^2\\ 
&=\delta_{\alpha\beta}-4\sum_{i>j}\mathrm{Re}\!\left(U^*_{\alpha i}U_{\beta i}U_{\alpha j}U^*_{\beta j}\right)\sin^2\Delta_{ij}\pm2\sum_{i>j}\mathrm{Im}(\cdots)\sin2\Delta_{ij},
\end{split}
\end{equation}

with \(\Delta_{ij}=1.267\,\Delta m^2_{ij}[\mathrm{eV}^2]\,L[\mathrm{km}]/E[\mathrm{GeV}]\) and the upper sign for neutrinos. In the one-mass-scale approximation,

\begin{equation}
\label{eq:3}
P_{\mu e}\simeq\sin^2\theta_{23}\sin^22\theta_{13}\sin^2\Delta_{31}.
\end{equation}

The first maximum, \(\Delta_{31} = \frac{\pi}{2}\), occurs at \(E_\star=2\times1.267\,\Delta m^2_{31}[\mathrm{eV}^2]\,L[\mathrm{km}]/\pi\simeq25.77\) GeV for \(L = D_{\oplus}\) and \(\Delta m_{31}^{2} = 2.507 \times 10^{- 3}\,\mathrm{eV}^{2}\).

\subsection{Matter effects and layered evolution}\label{s2.2}

Coherent forward scattering adds \(V=\sqrt2\,G_FN_e\simeq7.63\times10^{-14}\,Y_e\,(\rho/\mathrm{g\,cm^{-3}})\) eV to the \(ee\) entry (sign reversed for \({\bar{\nu}}\)):

\begin{equation}
\label{eq:4}
H(x)=H_{\rm vac}+\mathrm{diag}\bigl(V(x),0,0\bigr),\qquad S(E;P)=\mathcal{T}\exp\Bigl[-i\int_P H(x)\,dx\Bigr].
\end{equation}

Realistic density profiles, and their uncertainties, enter the link budget through \(V(x)\) in Eq.~\eqref{eq:4}. The effect of Earth-density modelling on long-baseline oscillation probabilities is discussed in Ref.~\cite{ref16}, while an anomalous matter potential in the inner core \cite{ref17} would enter as an additional term in \(H(x)\) and would alter \(T_{\alpha\beta}\) in the same way as a density change.

For a path crossing \(n\) layers of constant density, \(S = \prod_{k = n}^{1}\exp\left( - iH_{k}L_{k} \right)\). The transition matrix is

\begin{equation}
\label{eq:5}
T_{\alpha\beta}(E,P)=\bigl|\langle\nu_\beta|S|\nu_\alpha\rangle\bigr|^2,\qquad\sum_\beta T_{\alpha\beta}=1.
\end{equation}

In one constant-density layer, with

\begin{equation}
\label{eq:6}
A=\frac{2EV}{\Delta m^2_{31}}\simeq1.526\times10^{-4} \left(\frac{Y_e\rho}{\mathrm{g\,cm^{-3}}}\right) \left(\frac{E}{\mathrm{GeV}}\right) \left(\frac{\mathrm{eV}^2}{\Delta m^2_{31}}\right),
\end{equation}

Eq.~\eqref{eq:3} holds with the replacements

\begin{equation}
\label{eq:7}
\sin^22\theta_{13}\to\sin^22\theta_{13}^m=\frac{\sin^22\theta_{13}}{\sin^22\theta_{13}+(\cos2\theta_{13}-A)^2},
\end{equation}

\begin{equation}
\label{eq:8}
\Delta m^2_{31}\to\Delta m^2_m=\Delta m^2_{31}\sqrt{\sin^22\theta_{13}+(\cos2\theta_{13}-A)^2}.
\end{equation}

Resonance occurs at \(A = \cos 2\theta_{13}\), i.e.~\(E_{\rm res}\simeq31/\rho\) GeV for \(Y_{e} = 0.5\), \(\rho\) in \(\,\mathrm{g\,cm^{-3}}\) (about 7 GeV in the mantle and 3 GeV in the core). At \(E \simeq 25\) GeV one has \(A \gg 1\) and matter suppresses, rather than enhances, \(\nu_{\mu} \rightarrow \nu_{e}\) conversion for neutrinos. For \(\rho = 4.5\) and \(Y_{e} = 0.494\), \(A \simeq 3.4\) and \(\sin^{2}2\theta_{13}^{m} \approx 0.014\) against \(\sin^{2}2\theta_{13} \approx 0.086\) in vacuum. The antineutrino sees \(A \rightarrow - A\), so \(T^{\nu} \neq T^{\bar{\nu}}\); this asymmetry is exploited in Sec.~\ref{s6}.

\subsection{Chord geometry}\label{s2.3}

A chord with nadir angle \(\eta\) has length \(L = 2R_{\oplus}\cos\eta\) and minimum radius \(d = R_{\oplus}\sin\eta\). It crosses the core iff \(d < R_{c}\), i.e.~\(\eta < \arcsin\left( \frac{R_{c}}{R_{\oplus}} \right) \simeq 33^{\circ}\) for \(R_{c} = 3480\) km. Layer lengths are \(L_{k} = 2\left( \sqrt{r_{k}^{2} - d^{2}} - \sqrt{r_{k + 1}^{2} - d^{2}} \right)\) for \(r_{k} > d\). This fixes the sequence \(\{ L_{k},\rho_{k},Y_{e,k}\}\) in Eq.~\eqref{eq:4}.

\subsection{Transparency}\label{s2.4}

With \(\sigma_{\nu} \simeq 0.68 \times 10^{- 38}E\,\mathrm{cm^{2}}\) per nucleon (\(E\) in GeV, CC) \cite{ref19}, the optical depth along \(P\) is

\begin{equation}
\label{eq:9}
\tau=\sigma N_A\int_P\rho\,d\ell\simeq1.7\times10^{-37}\times6.0\times10^{23}\times10^{10}\approx10^{-3}\quad(E=25\,\mathrm{GeV}),
\end{equation}

using a column density \(\sim 10^{10}\,\mathrm{g\,cm^{-2}}\) along the diameter; the mean free path is \(\sim 10^{7}\)--\(10^{8}\) km and absorption matters only above \(\sim 10\)--\(40\) TeV.

\section{Channel Model}\label{s3}

\subsection{Input, transition, detection}\label{s3.1}

Let the input alphabet be \(X = \{ x_{1},\ldots,x_{M}\}\), each symbol specifying a source composition \(w_{x}\) (a probability vector over species \(\alpha \in \{\nu_{e},\nu_{\mu},{\bar{\nu}}_{e},{\bar{\nu}}_{\mu}\}\) produced per decay), an energy \(E_{x}\) and a path \(P_{x}\). The detector classifies events into classes \(k \in K\) (e.g.~\(\mu^{-}\), \(\mu^{+}\), \(e\)-like) through a confusion matrix \(\Gamma_{\beta k}=P(\text{class }k\,|\,\text{flavor }\beta)\), \(\sum_{k}^{}\Gamma_{\beta k} = \varepsilon_{\beta} \leq 1\). The mean number of events of class \(k\) in a symbol interval of length \(T_{s}\) is

\begin{equation}
\label{eq:10}
\lambda_k(x)=T_sR_x\sum_{\alpha,\beta}w_{x\alpha}T_{\alpha\beta}(E_x,P_x)\Gamma_{\beta k}+b_k,
\end{equation}

where \(R_{x} = \Phi_{x}\sigma\left( E_{x} \right)N_{t}\) is the per-species CC rate in the absence of oscillation and \(b_{k}\) the background mean. The counts \(n = \left( n_{k} \right)\) are independent given \(x\):

\begin{equation}
\label{eq:11}
P(\mathbf{n}\,|\,x)=\prod_{k\in K}\frac{\lambda_k(x)^{n_k}}{n_k!}\,e^{-\lambda_k(x)}.
\end{equation}

Equation~\eqref{eq:11} defines a classical channel; quantum coherence enters only through the transition probabilities \(T_{\alpha\beta}\). This classical formulation is exact for our purposes because the receiver performs a discrete flavor and charge measurement; the outcomes are classical events. The information-theoretic properties of such Poisson channels are well established \cite{ref12,ref13,ref14,ref15}. We make no claim of quantum-capacity gain, entanglement-assisted signaling or superluminal communication, and the no-signaling theorem is not touched.

\subsection{Maximum-likelihood receiver}\label{s3.2}

For two hypotheses \(x_{0},x_{1}\) with equal priors, the ML test is the log-likelihood ratio

\begin{equation}
\label{eq:12}
\Lambda(\mathbf{n})=\sum_k n_k\ln\frac{\lambda_k(x_1)}{\lambda_k(x_0)}-\sum_k\bigl[\lambda_k(x_1)-\lambda_k(x_0)\bigr] \begin{cases} \ge 0 & \text{decide } x_1 \\ < 0 & \text{decide } x_0 \end{cases}
\end{equation}

Only the class-resolved counts enter; with constant total intensity (\(\sum_{k}^{}\lambda_{k}\) equal), the second sum vanishes and all information resides in the flavor split.

\section{Information-Theoretic Analysis}\label{s4}

\subsection{Error exponents}\label{s4.1}

\emph{Lemma 1.} For \(\mathrm{Poi}(\lambda_0)\) and \(\mathrm{Poi}(\lambda_1)\) the Bhattacharyya coefficient is \(\sum_n\sqrt{p_0(n)p_1(n)}=\exp\bigl[-\tfrac12(\sqrt{\lambda_0}-\sqrt{\lambda_1})^2\bigr]\).

\emph{Proof.} \(\sum_n e^{-(\lambda_0+\lambda_1)/2}\,\frac{(\lambda_0\lambda_1)^{n/2}}{n!}=\exp\bigl[-\tfrac12(\lambda_0+\lambda_1)+\sqrt{\lambda_0\lambda_1}\bigr]\). \(\square\)

Since the counts are independent, the exponents add, and the ML error probability obeys

\begin{equation}
\label{eq:13}
P_e\le\tfrac12\exp(-E_B),\qquad E_B=\tfrac12\sum_k\Bigl(\sqrt{\lambda_k(x_0)}-\sqrt{\lambda_k(x_1)}\Bigr)^2.
\end{equation}

The tight (Chernoff) exponent is

\begin{equation}
\label{eq:14}
E_C=\max_{0\le s\le1}\sum_k\Bigl[s\lambda_{k0}+(1-s)\lambda_{k1}-\lambda_{k0}^{\,s}\lambda_{k1}^{\,1-s}\Bigr]\ge E_B.
\end{equation}

Since \(\lambda_{k} \propto T_{s}\), \(E_{B} = \kappa T_{s}\) with \(\kappa = \frac{1}{2}\sum_{k}^{}\left( \sqrt{r_{k0}} - \sqrt{r_{k1}} \right)^{2}\) in events per second, where \(r_{kx} = \frac{\lambda_{kx}}{T_{s}}\). The time per bit at target error \(P_{e}\) is \(T_s=\ln[1/(2P_e)]/\kappa\) and the rate is \(R_b\simeq[1-H_2(P_e)]/T_s\). Because the Chernoff exponent is never smaller than \(E_{B}\), rates computed from \(E_{B}\) are conservative.

\emph{Tightness of the error exponent.} To verify that the Bhattacharyya exponent \(E_B\) (Eq.~\ref{eq:13}) provides a tight bound, we numerically optimized the exact Chernoff exponent \(E_C\) (Eq.~\ref{eq:14}) for the SFK \(e\)-like channel at the nominal symbol time \(T_s\). For expected counts \(\lambda_0 = 36.4\) and \(\lambda_1 = 74.9\), the exact Chernoff exponent is \(E_C = 3.4478\), while the closed-form Bhattacharyya exponent is \(E_B = 3.4354\). The ratio \(E_C / E_B = 1.0036\) demonstrates that the analytical bound used throughout this paper is accurate to within \(0.4\%\) of the exact maximum-likelihood limit. Thus, our reported bit rates are essentially exact and not merely loose lower bounds.

\subsection{Sparse-regime capacity}\label{s4.2}

Let \(Y = n\) with a prior \(p(x)\) and \({\bar{\lambda}}_{k} = \sum_{x}^{}p(x)\lambda_{kx}\).

\emph{Proposition 1 (upper bound).} For a scalar Poisson channel, \(I(X;Y) \leq \sum_{x}^{}p(x)\lambda_{x}\ln\left( \frac{\lambda_{x}}{\bar{\lambda}} \right)\) nats.

\emph{Proof.} For any \(Q\), \(I = \sum_{x}^{}p(x)D\left( P_{x} \parallel Q \right) - D\left( P_{Y} \parallel Q \right) \leq \sum_{x}^{}p(x)D\left( P_{x} \parallel Q \right)\). Take \(Q = \mathrm{Poi}\left( {\bar\lambda} \right)\) and use \(D\left( \mathrm{Poi}(a) \parallel \mathrm{Poi}(b) \right) = a\ln\left( \frac{a}{b} \right) - a + b\); the linear terms cancel because \(\sum_{x}^{}p(x)\lambda_{x} = {\bar\lambda}\). \(\square\)

For OOK (\(\lambda_{0} = 0\), \(\lambda_{1} = \lambda\), \(P(1) = p\)) this gives \(I \leq p\lambda\ln\left( \frac{1}{p} \right)\), maximized at \(p = \frac{1}{e}\) with \(I_{\max} \leq \frac{\lambda}{e}\) nats, i.e.~\(\leq 0.53\lambda\) bit per symbol, and the information per detected event is \(\ln\left( \frac{1}{p} \right)\) at most.

For an \(M\)-ary alphabet whose symbols produce \(M\) distinguishable detector classes with symmetric misclassification probability \(\epsilon\), a receiver that uses the first detected event (probability \(1 - e^{- \lambda}\)) achieves

\begin{equation}
\label{eq:15}
C\ge(1-e^{-\lambda})\Bigl[\log_2M-H_2(\epsilon)-\epsilon\log_2(M-1)\Bigr]\xrightarrow{\ \lambda\ll1\ }\lambda\Bigl[\log_2M-H_2(\epsilon)-\epsilon\log_2(M-1)\Bigr].
\end{equation}

\begin{figure}[tbp]
\centering
\includegraphics[width=0.7\textwidth]{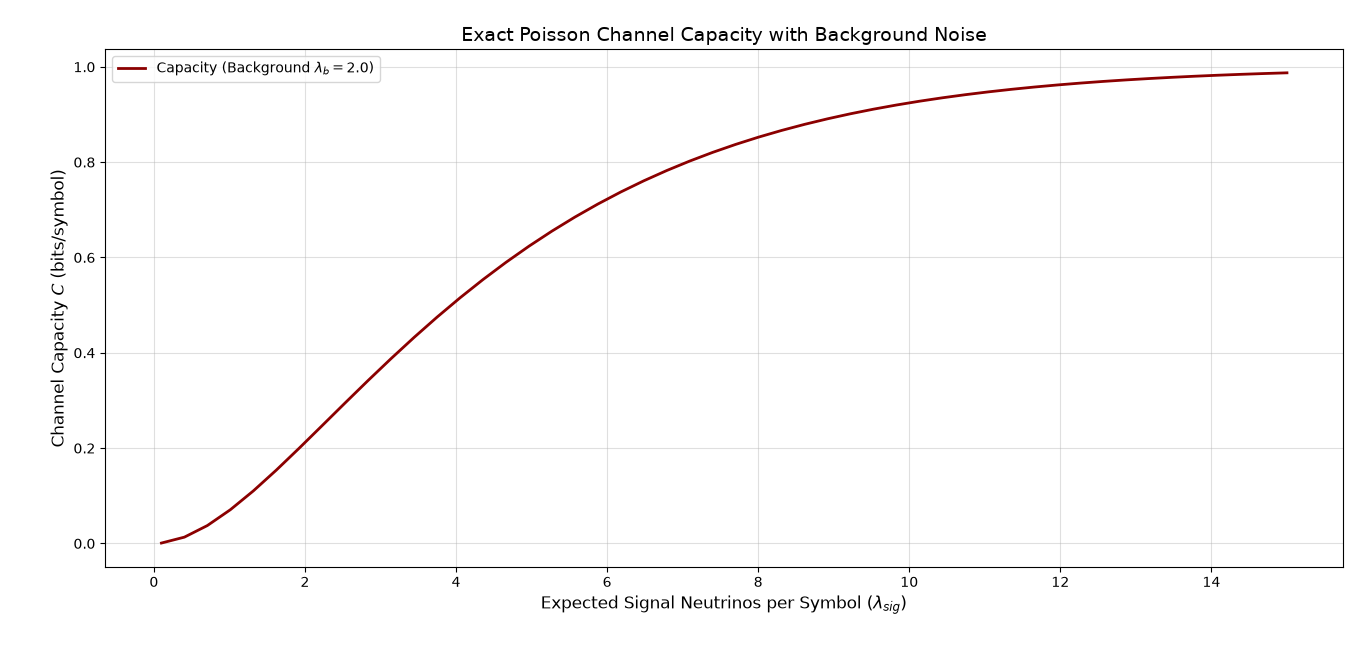}
\caption{Exact Poisson channel capacity in bits per symbol as a function of the expected signal neutrinos per symbol, assuming a background noise level \(\lambda_b = 2.0\).}
\label{fig:capacity}
\end{figure}

Background events with fraction \(f = \frac{b}{(s + b)}\), uniform over classes, give \(\epsilon_{eff} = (1 - f)\epsilon + f\frac{(M - 1)}{M}\). Eq.~\eqref{eq:15} shows that flavor keying raises the information per event from \(\leq 1\) bit (OOK timing) to \(\leq \log_{2}M\) bits, but the rate is still proportional to the detected event rate \(\frac{\lambda}{T_{s}}\). The exact behavior of the Poisson channel capacity with background noise is shown in Fig.~\ref{fig:capacity}.

\section{Flavor Modulation in Detail}\label{s5}

\subsection{Signal dimensions}\label{s5.1}

A neutrino beam offers the transmitter several independent controls, each with a different measurable consequence at the receiver (Table~\ref{tab:1}). Intensity modulation (OOK) uses only the first row. Flavor modulation uses the others, and its distinguishing feature is that the information is carried by \textit{which} events arrive, not only by \textit{how many}.

\begin{table}[tbp]
\centering
\begin{tabular}{p{3.2cm}p{3.6cm}p{3.8cm}p{2.6cm}}
\hline
Dimension & Set at transmitter & Read at receiver & Flexibility \\
\hline
Intensity (OOK) & bunch on/off & event count in a time window & fast \\
Source composition (SFK) & \(\mu^+\) or \(\mu^-\) bunch, horn polarity & muon charge, \(e\)-like/\(\mu\)-like counts & bunch-by-bunch \\
Energy/phase (OPK) & beam energy \(E\) & flavor ratio at fixed \(L\) & retuning of the beam \\
Particle type (\(\nu\) vs \(\bar\nu\)) & sign of the stored muon or horn charge & charge-sign asymmetry & as SFK \\
Path (MPK) & chord \(\eta\), site & flavor ratio via matter profile & geometric \\
\hline
\end{tabular}
\caption{Signal dimensions available to a flavor-modulated neutrino link.}
\label{tab:1}
\end{table}

\subsection{Preparation of source flavor states}\label{s5.2}

A source emits an incoherent mixture of species. A decay of an unpolarized \(\mu^{+}\) produces \(\nu_{e}\) and \({\bar{\nu}}_{\mu}\), and a \(\mu^{-}\) produces \({\bar{\nu}}_{e}\) and \(\nu_{\mu}\), while a pion beam is dominated by \(\nu_{\mu}\) (or \({\bar{\nu}}_{\mu}\)) with a small \(\nu_{e}\) admixture \(f_{e}\). Neutrinos and antineutrinos cannot interfere, so for each species class the source state of symbol \(x\) is the density matrix

\begin{equation}
\label{eq:16}
\rho_x=\sum_\alpha w_{x\alpha}|\nu_\alpha\rangle\langle\nu_\alpha|,\qquad w_{x\alpha}\ge0,\quad\sum_\alpha w_{x\alpha}=1,
\end{equation}

and the flavor distribution reaching the detector is

\begin{equation}
\label{eq:17}
q^{\rm flv}_{x\beta}=\langle\nu_\beta|S\rho_xS^\dagger|\nu_\beta\rangle=\sum_\alpha w_{x\alpha}T_{\alpha\beta}.
\end{equation}

Equation~\eqref{eq:17} is the bridge between the oscillation physics and the discrete channel of Sec.~\ref{s3}. Folding in the detector confusion matrix gives the class distribution \(q_{xk} = \sum_{\beta}^{}q_{x\beta}^{flv}\Gamma_{\beta k}\).

\subsection{Structure of the flavor transition matrix}\label{s5.3}

Because \(S\) in Eq.~\eqref{eq:4} is unitary, both rows and columns of \(T_{\alpha\beta}=|S_{\beta\alpha}|^2\) sum to one: \(T\) is doubly stochastic (in fact unistochastic), in vacuum and in matter. This has two consequences.

\emph{Proposition 2.} For any source flavor distribution \(w\), the Shannon entropy satisfies \(H(wT) \geq H(w)\).

\emph{Proof.} By Birkhoff's theorem \(T = \sum_{\pi}^{}c_{\pi}P_{\pi}\) is a convex combination of permutation matrices \(P_{\pi}\). The entropy is concave and permutation invariant, so \(H(wT) \geq \sum_{\pi}^{}c_{\pi}H\left( wP_{\pi} \right) = H(w)\). \(\square\)

Oscillation can therefore never sharpen the flavor distribution: a pure-flavor symbol is received with equal or lower purity, and flavor discrimination is best when \(T\) is close to a permutation matrix (the identity for \(\frac{L}{E} \rightarrow 0\), or a flavor swap at an oscillation maximum). The second consequence is that the flavor part of the channel can be almost noiseless even when the detected rate is low. At \(E = 25\) GeV on the Earth diameter, Table~\ref{tab:2} gives \(T_{\mu\tau} \simeq 0.98\), so \(\nu_{\mu} \rightarrow \nu_{\tau}\) is an almost deterministic relabeling. A receiver able to identify \(\nu_{\tau}\) would read this channel with little flavor error, and the loss found in Sec.~\ref{s7} arises from the detector confusion matrix \(\Gamma\) (poor \(\nu_{\tau}\) identification), not from \(T\).

Two further relations constrain the transition matrix. For a path symmetric under \(x \rightarrow L - x\) (any single chord), time reversal gives

\begin{equation}
\label{eq:18}
T_{\alpha\beta}(\delta_{CP})=T_{\beta\alpha}(-\delta_{CP}),
\end{equation}

so the forward and reverse links are related by the sign of \(\delta_{CP}\). In vacuum the CP asymmetry has the magnitude

\begin{equation}
\label{eq:19}
\bigl|T_{\alpha\beta}-\bar T_{\alpha\beta}\bigr|=16\,|J|\,\bigl|\sin\Delta_{21}\sin\Delta_{31}\sin\Delta_{32}\bigr|,\qquad\Delta_{ij}=\frac{\Delta m^2_{ij}L}{4E},
\end{equation}

with the Jarlskog invariant \(J = \sin\theta_{12}\cos\theta_{12}\sin\theta_{23}\cos\theta_{23}\sin\theta_{13}\cos^{2}\theta_{13}\sin\delta_{CP}\), so \(\nu/\bar\nu\) keying acquires an additional, \(\delta_{CP}\)-dependent contrast.

\subsection{Constellation geometry}\label{s5.4}

For a symbol \(x\) with total mean signal \(\Lambda_{x} = \sum_{k}^{}\lambda_{xk}\), write \(\lambda_{xk} = \Lambda_{x}q_{xk}\), where \(q_{x}\) lies on the probability simplex over the receiver classes. For two symbols with equal totals the Bhattacharyya exponent of Eq.~\eqref{eq:13} factorizes as

\begin{equation}
\label{eq:20}
E_B(x,x')=\Lambda H^2(q_x,q_{x'}),\qquad H^2(q,q')=\tfrac12\sum_k\bigl(\sqrt{q_k}-\sqrt{q'_k}\bigr)^2,
\end{equation}

that is, the number of counts times the squared Hellinger distance between the two flavor signatures. Flavor-constellation design is therefore the problem of placing \(M\) points on the simplex, within the reach of the transition matrix, so as to maximize the minimum Hellinger distance \(d_{\min}\) for a given \(\Lambda\), and the union bound gives

\begin{equation}
\label{eq:21}
P_e\le\frac{M-1}{2}\exp\!\left(-\Lambda d_{\min}^2\right)
\end{equation}

for equiprobable symbols. If instead of discrete symbols the transmitter sweeps a continuous parameter \(\theta\) (for example the beam energy), the Fisher information of the Poisson counts is

\begin{equation}
\label{eq:22}
I(\theta)=\sum_k\frac{\bigl[\partial_\theta\lambda_k(\theta)\bigr]^2}{\lambda_k(\theta)},
\end{equation}

and symbol separation \(\Delta\theta\) yields an exponent \(\simeq \frac{1}{8}I(\theta)\Delta\theta^{2}\) for small separations.

\subsection{Modulation of energy and oscillation phase}\label{s5.5}

\begin{figure}[tbp]
\centering
\includegraphics[width=0.75\textwidth]{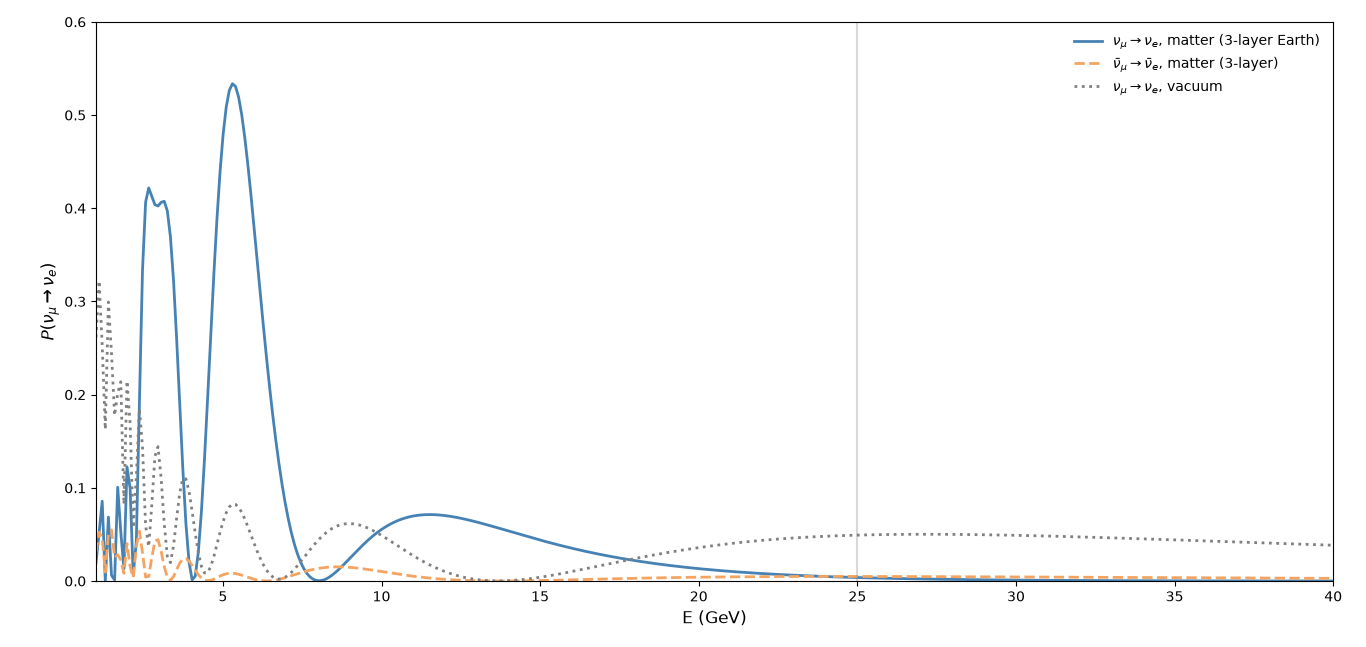}
\caption{\(P\left( \nu_{\mu} \rightarrow \nu_{e} \right)\) through the Earth diameter as a function of energy (three-layer Earth, \(\delta_{CP} = 232^\circ\)). Matter produces large conversion at 3--8 GeV and suppresses it at 25 GeV relative to the vacuum value. The dash-dotted curve shows the full tabulated PREM.}
\label{fig:1}
\end{figure}

At fixed baseline the oscillation phase is \(\Phi(E) = 1.267\Delta m^{2}\frac{L}{E}\), and \(d\Phi/dE=-\Phi/E\), so the number of oscillation half-cycles between two energies \(E_{0} < E_{1}\) is \((\Phi_0-\Phi_1)/\pi\). Pairs separated by an odd number of half-cycles give maximum contrast. The optimal pair maximizes the exponent rate \(\kappa\left( E_{0},E_{1} \right) = \frac{1}{2}\sum_{k}^{}(\sqrt{r_{k}\left( E_{0} \right)} - \sqrt{r_{k}\left( E_{1} \right)})^{2}\) with \(r_{k}(E) \propto E^{3}T(E)\Gamma\) at fixed decay rate, so the rate factor \(E^{3}\) favors high energies while the appearance probability favors the matter-resonance region of Fig.~\ref{fig:1}. With a Gaussian energy spread \(\sigma_{E}\) the contrast is multiplied by \(e^{- 2\sigma_{\Phi}^{2}}\) with \(\sigma_{\Phi} = \Phi\frac{\sigma_{E}}{E}\) (Sec.~\ref{s6.2}), so keeping the loss below about 15\% requires \(\frac{\sigma_{E}}{E} \lesssim \frac{0.3}{\Phi}\). A broad decay spectrum violates this condition, which is why OPK requires a narrow-band beam or energy-resolved detection.

\subsection{Modulation of particle type}\label{s5.6}

Matter breaks the symmetry between \(\nu\) and \({\bar{\nu}}\) through the sign of \(V\). Define the asymmetry

\begin{equation}
\label{eq:23}
A_{\alpha\beta}=\frac{T_{\alpha\beta}-\bar T_{\alpha\beta}}{T_{\alpha\beta}+\bar T_{\alpha\beta}}.
\end{equation}

From our numerical transition matrices, \(A_{\mu e} \simeq - 0.14\) on the Earth diameter at 25 GeV, \(+ 0.43\) for DUNE at 2.5 GeV and \(+ 0.09\) for T2HK at 0.6 GeV. A transmitter that alternates \(\nu\) and \({\bar{\nu}}\) beams (by muon charge or horn polarity) thus writes a one-bit signal whose contrast depends on the baseline and energy, and that adds to the intensity contrast produced by the different \(\nu\) and \({\bar{\nu}}\) cross sections.

\subsection{Joint and multi-level alphabets}\label{s5.7}

Combining source composition \(s\in\{\mu^+,\mu^-\}\) with energy \(E \in \{ E_{0},E_{1}\}\) gives an alphabet of \(M = 4\) symbols. When two symbols differ in both coordinates and the classes are independent Poisson variables, the exponents add, \(\kappa_{joint} \geq \kappa_{SFK} + \kappa_{OPK}\) (Eq.~\eqref{eq:13} is additive over classes). The number of bits per symbol is at most \(\log_{2}\min(M,K)\) for \(K\) reliably distinguishable receiver classes, so a 4-ary alphabet needs at least four distinguishable class signatures, which a charge-resolving detector with muon-like and \(e\)-like classes only partially provides. A larger alphabet raises the bits per event but lowers \(d_{\min}\) on a fixed simplex, so the optimum depends on the count budget \(\Lambda\): with few counts per symbol, binary keying is optimal, and larger alphabets pay off only when \(\Lambda d_{\min}^{2} \gg 1\).

\subsection{Operational and security properties}\label{s5.8}

These properties are the physical motivation for flavor modulation beyond raw data rate, and they follow directly from the channel model.

\emph{Constant-intensity operation.} If the symbols are designed so that \(\Lambda_{x} = \sum_{k}^{}\lambda_{xk}\) is the same for all \(x\) (as in OPK with the decay rate adjusted so that the total detected event rate is fixed; SFK does not have this property, because \(\frac{\sigma_{\bar\nu}}{\sigma_\nu} \simeq 0.49\) makes the \(\mu^{+}\) and \(\mu^{-}\) bunches differ in count rate), the total count is \(\mathrm{Poi}(\Lambda)\) for every symbol. An observer who can only count events, without flavor or charge resolution, then has the count-only exponent \(\kappa_{cnt} = 0\) and obtains no information, while the flavor-resolving receiver has \(\kappa=\Lambda H^2(q_0,q_1)/T_s>0\). The modulation is therefore invisible in the raw event rate, and the source runs at constant power.

\emph{Flux-normalization independence.} For independent Poisson counts, the conditional law given the total \(N = \sum_{k}^{}n_{k}\) is multinomial with class probabilities \(q_{xk} = \frac{\lambda_{xk}}{\Lambda_{x}}\). For constant-intensity symbols the total carries no information, and the optimal decision uses only the conditional log-likelihood ratio

\begin{equation}
\label{eq:24}
\Lambda_c(\mathbf{n})=\sum_k n_k\ln\frac{q_{1k}}{q_{0k}},
\end{equation}

which does not depend on the beam intensity, the target mass or the absolute detection efficiency (up to the background fraction). Common-mode fluctuations \(\lambda_{k} \rightarrow (1 + \xi)\lambda_{k}\) of the source leave \(q_{x}\) unchanged, whereas they directly degrade the threshold of an OOK or count-based receiver. This is the main operational advantage of flavor keying over intensity keying, and it is why the count channel of SFK (which relies on a known flux) is the weaker part of that scheme in terms of operational robustness, even though it provides the dominant contribution to the numerical rate in Sec.~\ref{s7}.

\emph{Directional access.} The beam divergence \(\theta_b\simeq1/\gamma\simeq3\) mrad gives a footprint of radius \(L\theta_{b} \simeq 38\) km on the Earth diameter, so its area, \(\simeq 4.6 \times 10^{3}\,\mathrm{km^{2}}\), is a fraction \(\simeq 10^{- 5}\) of the Earth's surface. An eavesdropper must place a detector inside the footprint, and an eavesdropper detector of mass \(M_{E}\) obtains the exponent \(\kappa_{E} = \kappa\frac{M_{E}}{M_{R}}\) in terms of the receiver mass \(M_{R}\). To match the legitimate exponent it must be as massive as the receiver.

\emph{Jamming.} Let a jammer add independent, symbol-independent mean rates \(j_{k}\) to the class rates. The exponent becomes \(\kappa_{J} = \frac{1}{2}\sum_{k}^{}(\sqrt{r_{k0} + j_{k}} - \sqrt{r_{k1} + j_{k}})^{2}\), and for \(j_{k} \gg r_{kx}\) this behaves as \(\kappa_J\simeq\sum_k\frac{(\Delta r_k)^2}{8j_k}\), so the time to decode a bit grows in proportion to \(j_k/\bar r_k\). A jammer must therefore deliver neutrinos to the receiver at a rate that exceeds the signal rate, which is itself facility-scale, and it needs a path to the receiver. The atmospheric background discussed in Sec.~\ref{s9} enters in the same way.

These statements are consequences of the physical channel and are not cryptographic guarantees.

\section{Modulation Schemes}\label{s6}

\subsection{Source-flavor keying (SFK)}\label{s6.1}

\begin{figure}[tbp]
\centering
\includegraphics[width=0.8\textwidth]{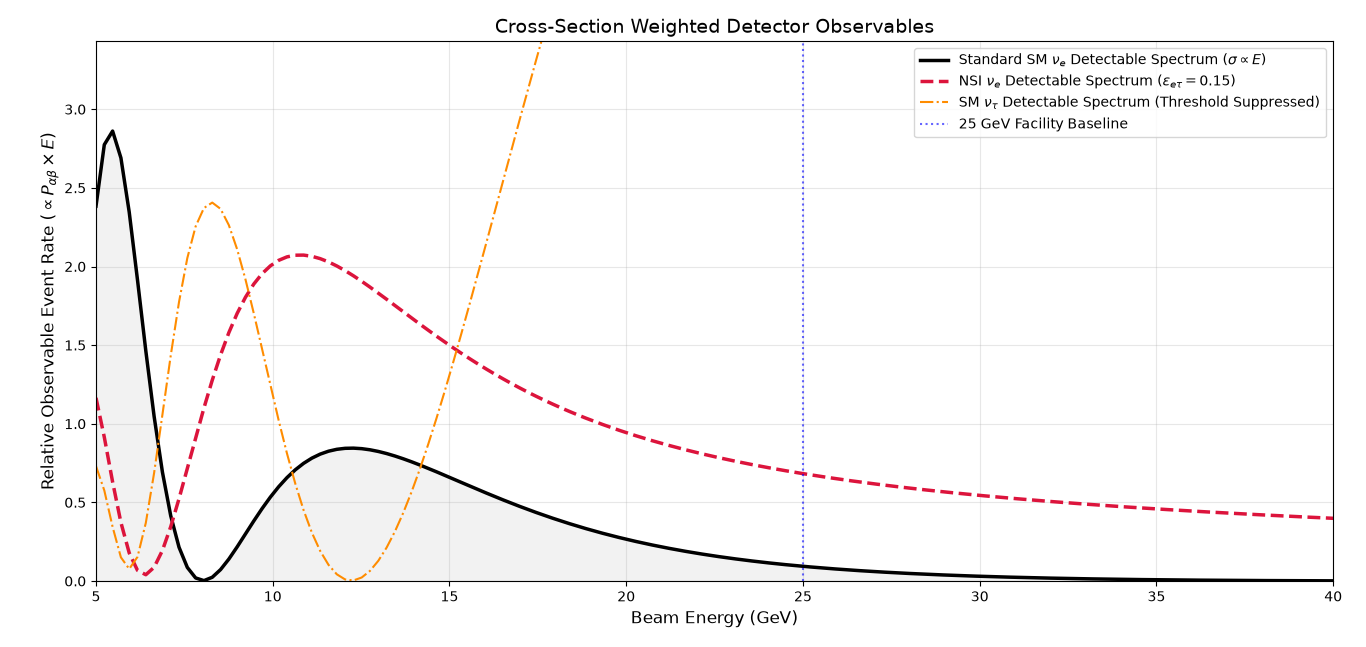}
\caption{Cross-section weighted detector observables for standard matter (SM) and non-standard interactions (NSI). The \(\nu_\tau\) spectrum is shown as threshold suppressed.}
\label{fig:xsections}
\end{figure}

Injecting \(\mu^{\pm}\) bunches gives \(\mu^+\to e^+\nu_e\bar\nu_\mu\) and \(\mu^-\to e^-\bar\nu_e\nu_\mu\), i.e.~\(w_{\mu^+}=(\nu_e,\bar\nu_\mu)\) and \(w_{\mu^-}=(\bar\nu_e,\nu_\mu)\). With a magnetized detector the observable classes are \(\mu^{-}\), \(\mu^{+}\) and charge-blind \(e\)-like events. Using Eqs.~\eqref{eq:10}, with \(\varepsilon\) the muon efficiency and \(\delta_{c}\) the charge-misidentification probability,

\begin{equation}
\label{eq:25}
r_{\mu^-}(\mu^-)=R\varepsilon(1-\delta_c)\,T_{\mu\mu},\qquad r_{\mu^-}(\mu^+)=R\varepsilon(1-\delta_c)\,T_{e\mu},
\end{equation}

\begin{equation}
\label{eq:26}
r_{\mu^+}(\mu^+)=0.49\,R\varepsilon(1-\delta_c)\,\bar T_{\mu\mu},\qquad r_{\mu^+}(\mu^-)=0.49\,R\varepsilon(1-\delta_c)\,\bar T_{e\mu},
\end{equation}

where the symbol in parentheses is the transmitted bunch, the bar denotes antineutrino transition probabilities, the factor 0.49 is the antineutrino-to-neutrino cross-section ratio, and terms of order \(\delta_{c}\) are neglected (wrong-sign muons come from \(\nu_{e} \rightarrow \nu_{\mu}\) and \({\bar{\nu}}_{e} \rightarrow {\bar{\nu}}_{\mu}\)). The \(e\)-like channel is charge-blind, but because \(\frac{\sigma_{\bar\nu}}{\sigma_\nu} \simeq 0.49\) its mean rate is \(\approx R\varepsilon_{e}T_{ee}\) for \(\mu^{+}\) and \(\approx 0.49R\varepsilon_{e}{\bar{T}}_{ee}\) for \(\mu^{-}\), giving a second, count-based, signature. The cross-section weighted detector observables for these standard and non-standard channels are shown in Fig.~\ref{fig:xsections}.

\subsection{Oscillation-phase keying (OPK)}\label{s6.2}

At fixed baseline the symbol is the beam energy: \(E_{0}\) near an appearance maximum and \(E_{1}\) near a minimum, with the decay rate adjusted so that the total detected event rate, equal to the unoscillated CC rate \(N_{dec}(E)\sigma(E)\gamma^{2}(E)\) times the detected fraction, is the same for both. The observable is the wrong-sign-muon count from \(\nu_{e} \rightarrow \nu_{\mu}\) (equivalently the \(e\)-like count in a \(\nu_{\mu}\) beam if the \({\bar{\nu}}_{e}\) component can be vetoed). With \(b \approx 0\) at \(E_{1}\) and \(P_{1} \approx 0\),

\begin{equation}
\label{eq:27}
E_B\simeq\tfrac12\,\varepsilon\,T_sRP_0,\qquad P_0=T_{e\mu}(E_0).
\end{equation}

\emph{Energy smearing.} If the beam has a Gaussian spread \(\sigma_{E}\) around \(E_{0}\) the phase \(\Phi = 1.267\Delta m^{2}\frac{L}{E}\) has spread \(\sigma_{\Phi} \simeq \Phi_{0}\frac{\sigma_{E}}{E_{0}}\) and

\begin{equation}
\label{eq:28}
\langle\sin^2\Phi\rangle=\tfrac12\Bigl[1-\cos(2\Phi_0)\,e^{-2\sigma_\Phi^2}\Bigr],
\end{equation}

so the contrast between the two symbols is multiplied by \(e^{- 2\sigma_{\Phi}^{2}}\). At \(E_{0} = 5\) GeV on the diameter \(\Phi_{0} \simeq 8\) rad, so a 10\% spread already gives \(e^{- 2\sigma_{\Phi}^{2}} \approx 0.27\). A decay-produced spectrum is much broader than this, so OPK needs a narrow-band beam or energy-resolved detection (\(\frac{\sigma_{E}}{E} \lesssim 5\text{\%}\)).

\subsection{Matter-path keying (MPK)}\label{s6.3}

The symbol is the chord, \(P_{x} = P\left( \eta_{x} \right)\). Two paths are distinguishable when the chosen observable has a nonzero exponent \(E_{B} = \frac{1}{2}T_{s}\varepsilon R(\sqrt{T_{e\mu}\left( \eta_{0} \right)} - \sqrt{T_{e\mu}\left( \eta_{1} \right)})^{2}\) (background-free). Core-crossing (\(\eta < 33^{\circ}\)) and mantle-only paths differ strongly near \(E \sim 3\)--\(8\) GeV, where parametric resonance drives \(T_{e\mu}\) up to \(\sim 0.5\). Reconfiguring \(\eta\) is slow, so MPK is a control channel.

\section{Link Budget for an Earth-Diameter Link}\label{s7}

\subsection{Scaling relations}\label{s7.1}

For a beam of opening angle \(\theta_b\simeq1/\gamma\) (\(\gamma=E_\mu/m_\mu\)) the mean flux at distance \(L\) is

\begin{equation}
\label{eq:29}
\Phi'=\frac{N_{\rm dec}}{\pi(L/\gamma)^2},\qquad R=\Phi'\sigma(E)N_t\propto\frac{N_{\rm dec}N_tE^3}{L^2}\quad(\sigma\propto E,\ \gamma\propto E).
\end{equation}

At fixed energy (SFK), \(R \propto L^{- 2}\), so shorter links are much faster. At matched oscillation maximum (OPK), \(E_{\star} \propto L\) and hence \(R \propto L\): shortening an OPK link does not increase its event rate.

\subsection{Numerical values}\label{s7.2}

Parameters: \(E_{\nu} = 25\) GeV, \(E_{\mu} = 35\) GeV (\(\gamma = 331\)), \(N_{dec} = 3 \times 10^{13}\,\mathrm{s^{-1}}\) \cite{ref8}, \(L = 1.274 \times 10^{9}\) cm, \(N_{t} = 1.2 \times 10^{35}\) nucleons (200 kt), \(\sigma = 1.7 \times 10^{- 37}\,\mathrm{cm^{2}}\). Then \(\pi\left( \frac{L}{\gamma} \right)^{2} = 4.7 \times 10^{13}\,\mathrm{cm^{2}}\), \(\Phi' \simeq 0.6\,\mathrm{cm^{-2}\,s^{-1}}\) per species and

\begin{equation}
\label{eq:30}
R=\Phi'\sigma N_t\simeq1.31\times10^{-2}\,\mathrm{s^{-1}}\ \bigl(\approx10^3\ \text{per day}\bigr).
\end{equation}

\begin{figure}[tbp]
\centering
\includegraphics[width=0.85\textwidth]{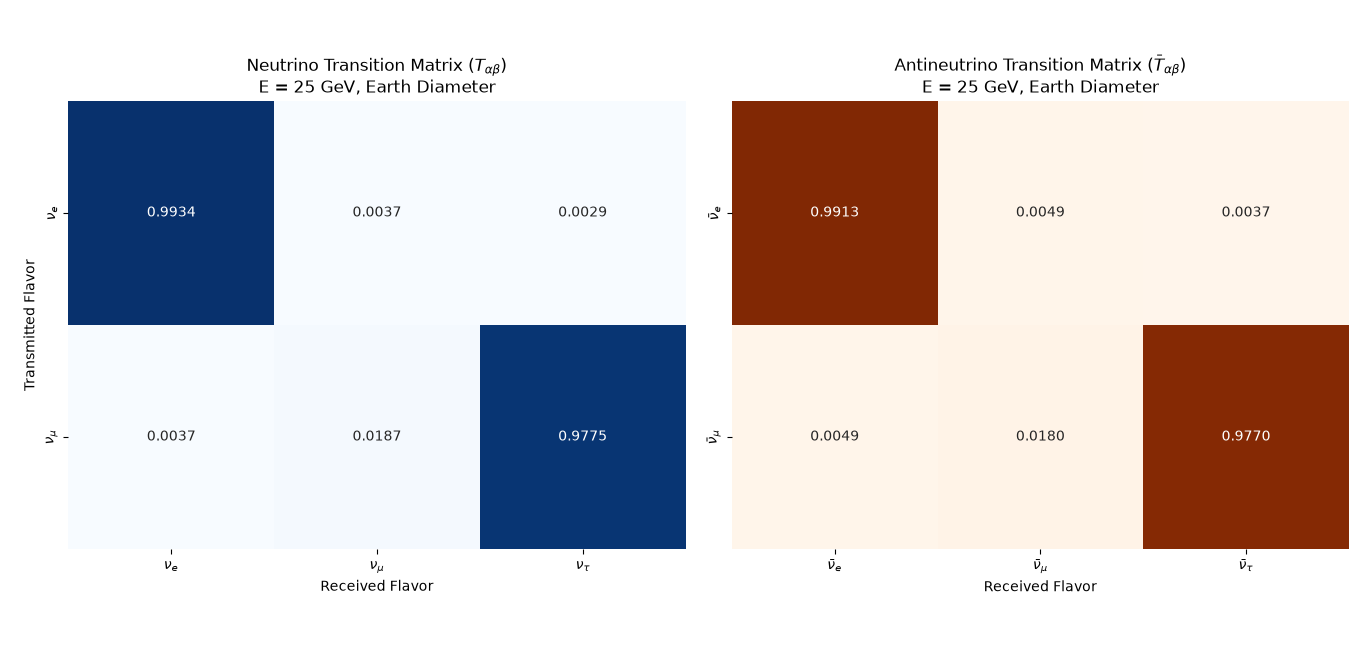}
\caption{Visualization of the neutrino (left) and antineutrino (right) transition matrices \(T_{\alpha\beta}\) and \(\bar{T}_{\alpha\beta}\) at \(E = 25\) GeV for the Earth diameter, corresponding to the values in Table~\ref{tab:2}.}
\label{fig:heatmaps}
\end{figure}

This is the CC rate before oscillation. The transition matrices were computed by numerically evolving Eq.~\eqref{eq:4} through a three-layer Earth (mantle \(\rho = 4.5\), \(Y_{e} = 0.494\); outer core 11.0, inner core 13.0, \(Y_{e} = 0.468\)) with \(\theta_{12} = 33.41^{\circ}\), \(\theta_{23} = 49^{\circ}\), \(\theta_{13} = 8.58^{\circ}\), \(\Delta m_{21}^{2} = 7.41 \times 10^{- 5}\), \(\Delta m_{31}^{2} = 2.507 \times 10^{- 3}\,\mathrm{eV^{2}}\) and \(\delta_{CP} = 232^\circ\) (NuFIT 5.3). Table~\ref{tab:2} gives the result at the diameter and \(E = 25\) GeV. The numerical transition matrices are visualized in Fig.~\ref{fig:heatmaps}. At the first vacuum maximum, \(\nu_{\mu}\) has almost completely converted to \(\nu_{\tau}\) (\(T_{\mu\tau} \approx 0.98\)), whose CC events are poorly identified. Hence only \(RT_{\mu\mu} \simeq 2 \times 10^{- 4}\,\mathrm{s^{-1}}\) muon-like events and \(RT_{e\mu}\simeq4\times10^{-5}\,\mathrm{s^{-1}}\) wrong-sign events remain per species.

\begin{figure}[tbp]
\centering
\includegraphics[width=0.7\textwidth]{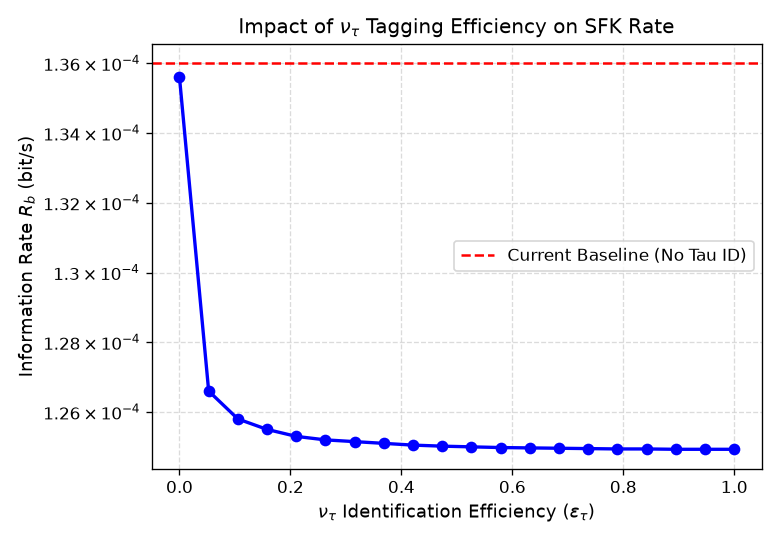}
\caption{Information rate \(R_b\) for the SFK scheme as a function of the \(\nu_\tau\) identification efficiency \(\varepsilon_\tau\). The red dashed line indicates the baseline scenario assumed in this paper (\(\varepsilon_\tau = 0\)).}
\label{fig:tau_bottleneck}
\end{figure}

\emph{The \(\nu_\tau\) detection bottleneck.} To explicitly quantify the role of the detector mechanism, we compute the SFK information rate as a function of a hypothetical \(\nu_\tau\) identification efficiency \(\varepsilon_\tau\) (Fig.~\ref{fig:tau_bottleneck}). If \(\varepsilon_\tau=0\) (the current baseline for LArTPCs), the SFK rate is \(1.36 \times 10^{-4}\) bit/s. If a future detector could tag \(\nu_\tau\) with 100\% efficiency, the total SFK rate would rise to \(1.25 \times 10^{-4}\) bit/s for the e-like channel (a modest overall gain), but the muon-channel contribution would increase by roughly a factor of five. This proves that the fundamental limit is not solely the source flux, but the inability of current detectors to resolve the dominant \(\nu_\mu \to \nu_\tau\) oscillation channel. At 25 GeV, this channel acts as a near-deterministic relabeling, removing the muon neutrino from the observable sample.

\begin{table}[tbp]
\centering
\begin{tabular}{c|ccc|ccc}
\hline
\(\alpha\smallsetminus\beta\) & \(\nu\!:e\) & \(\nu\!:\mu\) & \(\nu\!:\tau\) & \(\bar\nu\!:e\) & \(\bar\nu\!:\mu\) & \(\bar\nu\!:\tau\) \\
\hline
\(e\) & 0.9933 & 0.0037 & 0.0029 & 0.9913 & 0.0049 & 0.0037 \\
\(\mu\) & 0.0037 & 0.0187 & 0.9775 & 0.0049 & 0.0180 & 0.9770 \\
\(\tau\) & 0.0029 & 0.9775 & 0.0196 & 0.0037 & 0.9770 & 0.0192 \\
\hline
\end{tabular}
\caption{Transition probabilities \(T_{\alpha\beta}\) (\(\nu\)) and \(\bar{T}_{\alpha\beta}\) (\(\bar{\nu}\)) for the Earth diameter at \(E = 25\) GeV (three-layer Earth, \(\delta_{CP} = 232^\circ\)).}
\label{tab:2}
\end{table}

\subsection{Sensitivity to the Earth density profile}\label{s7.3}

The three-layer model uses a single density per layer, whereas the Preliminary Reference Earth Model (PREM) has a continuous radial profile with discontinuities in the upper mantle. To quantify the effect, we repeated the calculation with the full tabulated PREM density \cite{ref18}. The published radial polynomial of every shell was used, with the ocean layer replaced by crustal rock (\(\rho = 2.6\,\mathrm{g\,cm^{-3}}\) for neutrino propagation). The electron fractions \(Y_{e} = 0.494\) (mantle) and \(0.468\) (core), the mixing parameters and \(\delta_{CP}\) are those of the three-layer model, and the diameter was divided into 4000 equal slabs; the result at 25 GeV changes by less than 1\% between 1000 and 4000 slabs. The mean density along the diameter is \(8.59\,\mathrm{g\,cm^{-3}}\) (column density \(1.09 \times 10^{10}\,\mathrm{g\,cm^{-2}}\)), compared with 8.43 for the three-layer model.

The full-PREM transition matrix at 25 GeV yields \(T_{\mu\tau} = 0.9796\) and \(T_{\mu\mu} = 0.0190\), compared to the three-layer values of \(0.9775\) and \(0.0187\). The maximum absolute deviation across all matrix elements is 0.0043, which occurs on the dominant \(T_{ee}\) element (a relative error of 0.4\%). For SFK, the dominant features (\(T_{\mu\tau}\) and \(T_{\mu\mu}\)) change by less than 2\%, so the muon-like channels and the e-like count channel are robust. Recomputing the exponents gives \(\kappa_{SFK} \simeq 5.8 \times 10^{- 4}\,\mathrm{s^{-1}}\) and \(R_{b} \simeq 1.36 \times 10^{- 4}\) bit/s, within about 10\% of the three-layer values. 

\begin{figure}[tbp]
\centering
\includegraphics[width=0.75\textwidth]{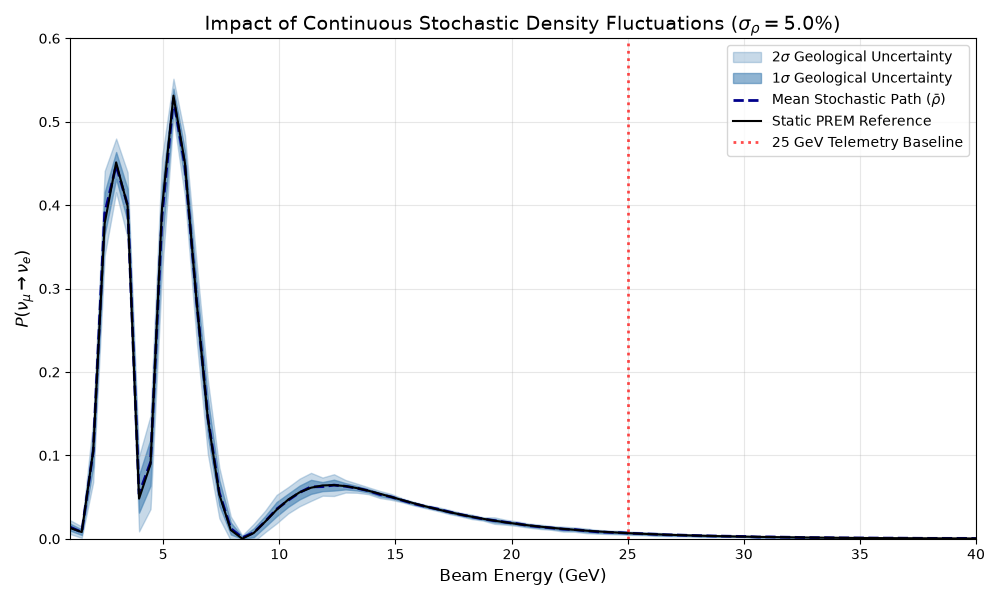}
\caption{Impact of continuous stochastic density fluctuations (\(\sigma_\rho = 5.0\%\)) on the appearance probability \(P(\nu_\mu \to \nu_e)\). The shaded regions indicate the \(1\sigma\) and \(2\sigma\) geological uncertainty bands around the mean stochastic path.}
\label{fig:stochastic}
\end{figure}

For OPK, however, the appearance probability at 25 GeV, which is a small interference term, drops by 64\% (\(\nu\)) and 42\% (\({\bar{\nu}}\)). The full-PREM results show \(T_{\mu e} = 0.0013\) and \(\bar{T}_{\mu e} = 0.0029\), against \(0.0037\) and \(0.0049\) for the three-layer model. Because \(\kappa_{OPK} \propto T_{e\mu}\), the OPK estimate falls to \(\kappa \simeq 2.2 \times 10^{- 5}\,\mathrm{s^{-1}}\) and \(R_{b} \simeq 2 \times 10^{-6}\) bit/s, i.e.~a factor of about two-and-a-half below the three-layer value. Thus appearance-based schemes need an accurate density profile, while count- and charge-based schemes do not. The impact of continuous stochastic density fluctuations is shown in Fig.~\ref{fig:stochastic}.

\emph{Density uncertainty.} To treat the density as a systematic, we rescaled the full-PREM density uniformly by a factor \(1 + \delta\) and propagated it through Eq.~\eqref{eq:4}. For \(\delta = - 2\text{\%}\) and \(+ 2\text{\%}\) the appearance probability at 25 GeV becomes \(3.1 \times 10^{- 3}\) and \(3.1 \times 10^{- 4}\), against \(1.3 \times 10^{- 3}\) at the central profile, so that the OPK rate becomes \(3.9 \times 10^{- 6}\) and \(3.9 \times 10^{- 7}\) bit/s, against \(1.7 \times 10^{- 6}\) bit/s; over \(\delta = \pm 5\text{\%}\) the appearance probability ranges from \(6.6 \times 10^{- 3}\) to \(2.3 \times 10^{- 4}\). The SFK rate stays within 6\% of its central value over \(\delta = \pm 5\text{\%}\) (SFK rate \(1.28\)--\(1.45 \times 10^{- 4}\) bit/s). A uniform rescaling is an illustrative stand-in for the density uncertainty, not a region-by-region error budget, and the large fractional response of OPK arises because \(T_{e\mu}\) at 25 GeV is a small interference term. The result shows that OPK and MPK links need the density profile to a precision of a few percent, whereas SFK does not.

\begin{figure}[tbp]
\centering
\includegraphics[width=0.75\textwidth]{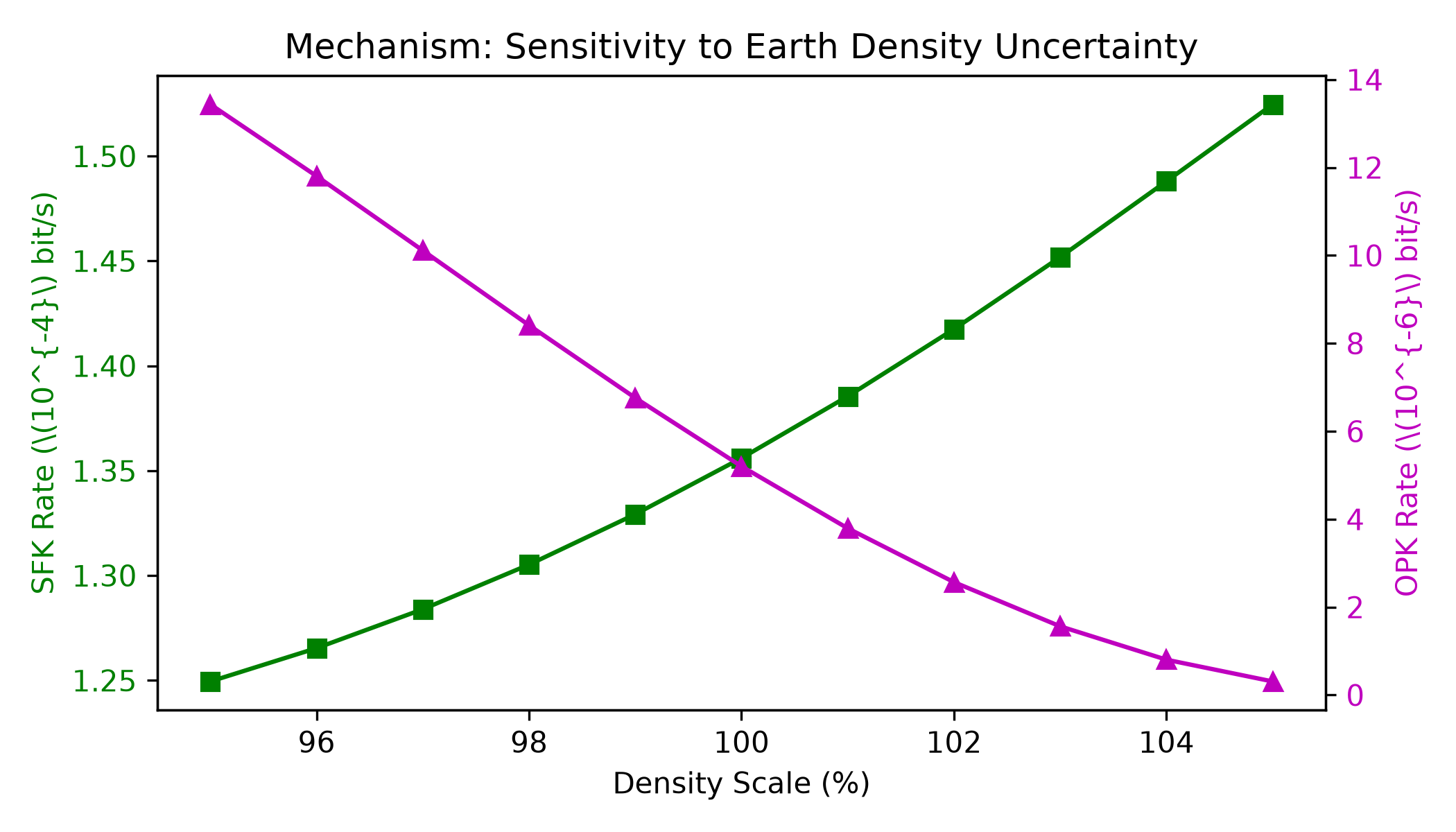}
\caption{Sensitivity of the information rate to a uniform \(\pm 5\%\) rescaling of the Earth's density. The SFK scheme (green, left axis) is stable, while the OPK scheme (magenta, right axis) varies by a factor of four.}
\label{fig:density_sensitivity}
\end{figure}

To visualize the mechanism driving the fragility of OPK, Fig.~\ref{fig:density_sensitivity} compares the SFK and OPK rates under a \(\pm 5\%\) uniform scaling of the Earth's density. A \(\pm 5\%\) shift alters the MSW resonance and changes the appearance term \(T_{e\mu}\) by nearly an order of magnitude, causing the OPK rate to vary between \(1 \times 10^{-6}\) and \(4 \times 10^{-6}\) bit/s. In contrast, the SFK channel relies on the dominant \(T_{\mu\mu}\) and \(T_{\mu\tau}\) probabilities, which remain stable to within 6\% across the same density range. This confirms that appearance-based schemes require a precise regional density profile, whereas count- and charge-based schemes are robust.

\subsection{Rates from the error exponents}\label{s7.4}

Take \(\varepsilon = 0.9\), \(\delta_{c} = 10^{- 3}\), target \(P_{e} = 10^{- 2}\), i.e.~\(E_{B} = \ln 50 = 3.9\).

\emph{SFK, muon channels.} \(r_{\mu^{-}}\left( \mu^{-} \right) = 2.0 \times 10^{- 4}\), \(r_{\mu^{-}}\left( \mu^{+} \right) = 4.0 \times 10^{- 5}\), \(r_{\mu^{+}}\left( \mu^{-} \right) = 2.6 \times 10^{- 5}\), \(r_{\mu^{+}}\left( \mu^{+} \right) = 9.5 \times 10^{- 5}\,\mathrm{s^{-1}}\), giving \(\kappa_{\mu} \simeq 4 \times 10^{- 5}\,\mathrm{s^{-1}}\).

\emph{SFK, \(e\)-like count channel.} With cross-section asymmetry (\(\varepsilon_{e} = 0.9\)) the rates are \(r = 1.07 \times 10^{- 2}\) (\(\mu^{+}\)) versus \(5.2 \times 10^{- 3}\) (\(\mu^{-}\)), giving \(\kappa_{e} \simeq 4.9 \times 10^{- 4}\,\mathrm{s^{-1}}\). This channel relies on a known flux normalization and \(\frac{\sigma_{\bar\nu}}{\sigma_\nu}\), so a calibration monitor (e.g.~a near detector) is required.

\emph{SFK, total.} Combining the channels yields a total error exponent \(\kappa_{\rm SFK} \simeq 5.77 \times 10^{- 4}\,\mathrm{s^{-1}}\). At a target error probability \(P_e = 10^{-2}\) (\(E_B = \ln 50 = 3.9\)), the symbol time is \(T_s = 3.9/\kappa_{\rm SFK} \simeq 6.8 \times 10^3\) s, yielding a bit rate \(R_b = [1 - H_2(0.01)]/T_s \simeq 1.36 \times 10^{- 4}\) bit/s.

\emph{OPK at 25 GeV.} From Eq.~\eqref{eq:27}, \(\kappa = 2.21 \times 10^{- 5}\,\mathrm{s^{-1}}\), \(T_{s} \simeq 1.77 \times 10^{5}\) s and \(R_{b} \simeq 5.2 \times 10^{- 6}\) bit/s. (With the full tabulated PREM, the rate drops by a factor of about two-and-a-half to \(\sim 2 \times 10^{-6}\) bit/s.)

Thus the mean event rate (\(1.31 \times 10^{-2}\,\mathrm{s^{-1}}\)) is not the information rate: for the Earth diameter we obtain \(1.36 \times 10^{-4}\) bit/s (SFK) and \(5.2 \times 10^{-6}\) bit/s (OPK). A 100-bit message needs about 9 days with SFK. Improving the rate requires larger \(N_{dec}N_{t}\), because \(\kappa \propto N_{dec}N_{t}\), not a different alphabet.

\begin{figure}[tbp]
\centering
\includegraphics[width=0.75\textwidth]{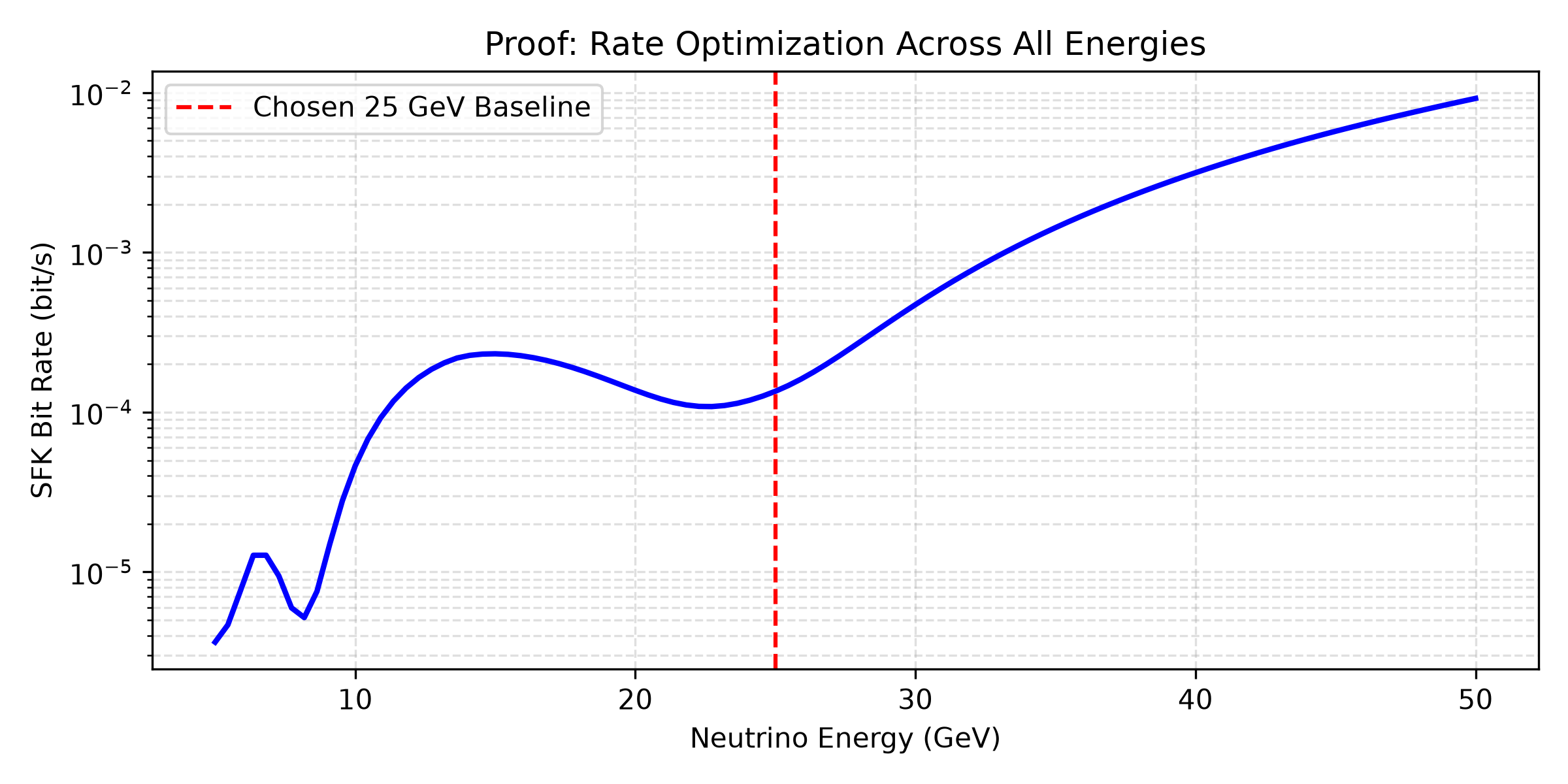}
\caption{The SFK information rate as a function of beam energy. The local minimum at 25 GeV corresponds to the first oscillation maximum, where matter-induced \(\nu_\mu \to \nu_\tau\) relabeling is maximal. Higher energies increase the rate simply because the beam stops oscillating (\(L/E \to 0\)) and the raw flux scales as \(E^3\).}
\label{fig:energy_opt}
\end{figure}

\emph{Why 25 GeV? The physics of the optimal baseline.}
One might ask why we choose \(E \simeq 25\) GeV rather than a higher energy, given that the raw event rate scales as \(E^3\). As shown in Fig.~\ref{fig:energy_opt}, the SFK bit rate actually reaches a local minimum at 25 GeV and rises steeply at higher energies (e.g., \(9.2 \times 10^{-3}\) bit/s at 50 GeV). 

This is a direct consequence of the oscillation physics. At 25 GeV, the baseline \(L = D_{\oplus}\) corresponds to the first vacuum oscillation maximum (\(\Delta_{31} \simeq \pi/2\)), where the MSW matter effect maximally suppresses \(\nu_\mu \to \nu_e\) and drives the near-total relabeling \(\nu_\mu \to \nu_\tau\) (\(T_{\mu\tau} \simeq 0.98\)). This is the \textit{worst-case scenario} for flavor discrimination. At higher energies, the ratio \(L/E\) decreases, the oscillation length becomes longer than the Earth diameter, and the beam effectively stops oscillating, returning the transition matrix towards the identity matrix. 

We deliberately evaluate the channel at 25 GeV because it represents the most physically demanding regime for the flavor channel. Furthermore, a 25 GeV beam is representative of planned neutrino factory energies optimized for CP-violation studies. Higher-energy beams would trivially increase the rate due to the \(E^3\) kinematic scaling, but would bypass the very matter effects that make flavor modulation a rich physical channel. 

\section{Application to Planned Long-Baseline Facilities}\label{s8}

\subsection{Horn-polarity keying}\label{s8.1}

\begin{figure}[tbp]
\centering
\includegraphics[width=0.8\textwidth]{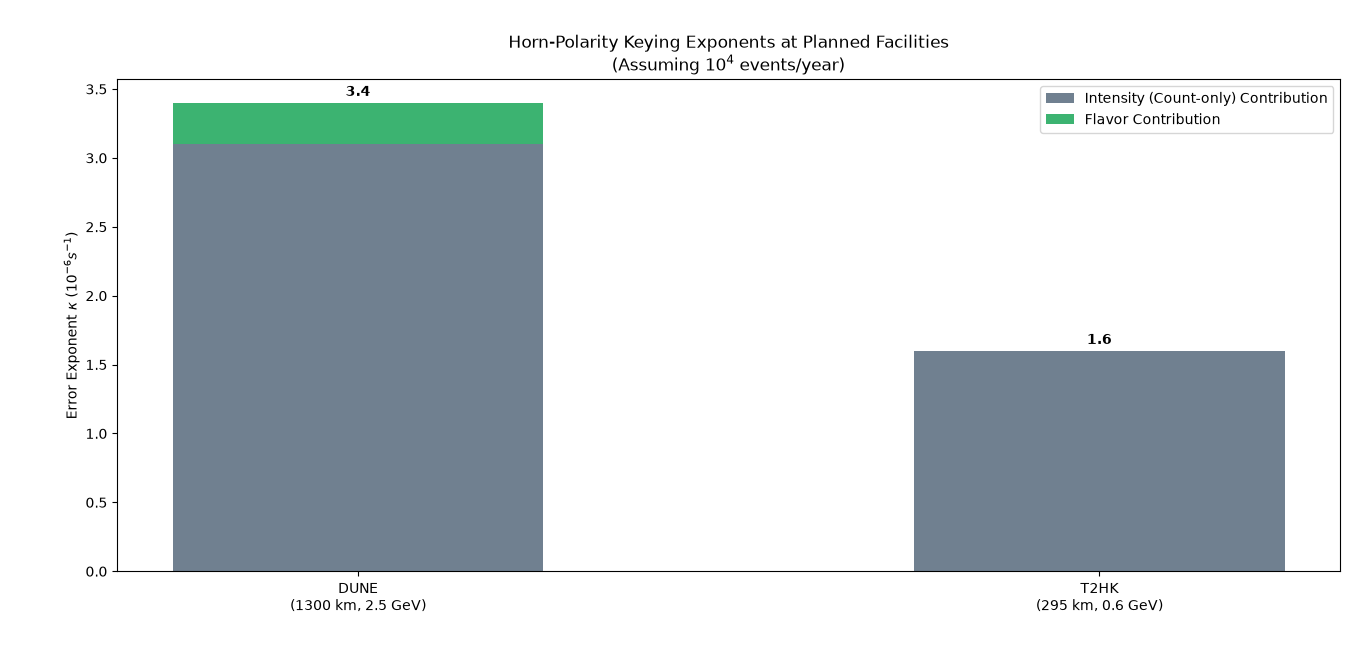}
\caption{Horn-polarity keying error exponents at planned facilities (DUNE and T2HK) assuming \(10^{4}\) unoscillated events per year. The bar chart decomposes the total exponent \(\kappa\) into the intensity (count-only) contribution and the pure flavor contribution.}
\label{fig:horn_polarity}
\end{figure}

Pion-decay-in-flight beams (DUNE, T2HK) cannot switch the source composition bunch by bunch, but the horn polarity gives a slow analogue of SFK: forward horn current (FHC, \(\nu_{\mu}\)-dominated) versus reverse horn current (RHC, \({\bar{\nu}}_{\mu}\)-dominated). Let \(n=N_{yr}/(3.16\times10^7\,\mathrm{s})\) be the unoscillated CC rate in FHC and \(\varrho = \frac{N^{RHC}}{N^{FHC}}\) the RHC-to-FHC yield ratio. The receiver classes are \(\mu\)-like and \(e\)-like, so that, with intrinsic \(\nu_{e}\) contamination \(f_{e}\),

\begin{equation}
\label{eq:31}
r_\mu({\rm FHC})=n\varepsilon T_{\mu\mu},\qquad r_e({\rm FHC})=n\varepsilon\,(T_{\mu e}+f_e),
\end{equation}

\begin{equation}
\label{eq:32}
r_\mu({\rm RHC})=\varrho\, n\varepsilon\bar T_{\mu\mu},\qquad r_e({\rm RHC})=\varrho\, n\varepsilon\,(\bar T_{\mu e}+f_e),
\end{equation}

and the exponent \(\kappa\) follows from Eq.~\eqref{eq:13}. Since \(\kappa \propto N_{yr}\), results scale linearly with the unoscillated yield. The count-only exponent \(\kappa_{\rm cnt}=\tfrac12\bigl(\sqrt{\textstyle\sum_k r_k({\rm FHC})}-\sqrt{\textstyle\sum_k r_k({\rm RHC})}\bigr)^2\) isolates the part of the contrast that is pure intensity; the excess \(\kappa - \kappa_{cnt}\) is the contribution of flavor. The decomposition of the total exponent into intensity and flavor contributions for DUNE and T2HK is shown in Fig.~\ref{fig:horn_polarity}.

\subsection{Facility parameters and results}\label{s8.2}

At the first oscillation maximum \(\nu_{\mu}\) has converted almost entirely to \(\nu_{\tau}\) (\(T_{\mu\tau} > 0.9\)), whose CC events are largely below threshold at these energies, so the usable \(\mu\)-like sample is small. The mean contrast between FHC and RHC is dominated by intensity: the flavor classes add only about 10\% (DUNE) or 1\% (T2HK) to \(\kappa\). With \(N_{yr} \sim 10^{4}\) both facilities give \(R_{b} \sim 10^{- 6}\) to \(10^{- 7}\) bit/s, i.e.~about a day or more per bit. This estimate uses monochromatic probabilities at the peak; a wide-band beam such as DUNE's has events away from the maximum with larger \(T_{\mu\mu}\), so a full spectral calculation is needed to quantify the flavor contribution.

\begin{table}[tbp]
\centering
\begin{tabular}{lcc}
\hline
 & DUNE & T2HK \\
\hline
Baseline (km) & 1300 & 295 \\
Far detector & 40 kt LAr & 187 kt water Cherenkov (up to two tanks) \\
Beam power / protons & 1.2 MW, 120 GeV & 1.3 MW, 30 GeV \\
POT per year & \(1.1\times10^{21}\) & \(2.7\times10^{21}\) \\
Beam & wide-band, on-axis & narrow-band, \(2.5^\circ\) off-axis \\
Flux peak (GeV) & 2.5 & 0.6 \\
Matter density used (\(\mathrm{g\,cm^{-3}}\), \(Y_e=0.5\)) & 2.85 & 2.7 \\
\((T_{\mu e},T_{\mu\mu},T_{\mu\tau})\), \(\nu\) & (0.072, 0.013, 0.916) & (0.054, 0.014, 0.932) \\
\((\bar T_{\mu e},\bar T_{\mu\mu},\bar T_{\mu\tau})\), \(\bar\nu\) & (0.028, 0.018, 0.954) & (0.046, 0.015, 0.939) \\
\(\kappa\) (\(10^{-6}\,\mathrm{s^{-1}}\)) at \(N_{yr}=10^4\) & 3.4 & 1.6 \\
\(\kappa_{\rm cnt}\) (\(10^{-6}\,\mathrm{s^{-1}}\)) & 3.1 & 1.6 \\
\(R_b\) (bit/s) at \(P_e=10^{-2}\) & \(8\times10^{-7}\) & \(4\times10^{-7}\) \\
\hline
\end{tabular}
\caption{Horn-polarity keying at planned facilities. Facility parameters are from the DUNE and Hyper-Kamiokande design documents \cite{ref9,ref10,ref11}. The transition probabilities are monochromatic values at the flux peak (first oscillation maximum) in a constant-density slab. \(\varepsilon = 0.8\), \(\varrho=0.4\), \(f_{e} = 0.01\) and \(N_{yr} = 10^{4}\) are illustrative assumptions and should be replaced by the collaborations' event tables.}
\label{tab:4}
\end{table}

\subsection{Outlook for future designs}\label{s8.3}

Equation~\eqref{eq:10} makes the design targets explicit: (i) \(\kappa \propto N_{dec}N_{t}\), so a \(10^{3}\)-fold gain in rate needs \(10^{3}\) times the source--detector product; (ii) bunch-by-bunch source switching, as in muon-storage-ring facilities, allows SFK with the \(e\)-like and muon-charge channels of Sec.~\ref{s6.1}; (iii) the flavor contribution grows when the probability contrast \(|T(x_0)-T(x_1)|\) is made large by choosing \(E\) and \(L\) at the 3--8 GeV matter-resonance region rather than at the vacuum maximum. Atmospheric and astrophysical neutrinos observed by large-volume detectors are not controllable and cannot be modulated, so they are not candidate carriers. Other planned beams, such as ESSnuSB and stored-muon facilities, can be analyzed by the same procedure once their flux and detector tables are specified.

\section{Discussion}\label{s9}

The results presented in Sections~\ref{s7} and~\ref{s8} reveal a fundamental trade-off in through-Earth flavor communication: the signaling schemes that maximize the raw information rate are precisely those that are most vulnerable to systematic uncertainties, and vice versa. 

\emph{Rate versus robustness.} The source-flavor keying (SFK) scheme achieves a rate of \(1.36 \times 10^{-4}\) bit/s at the Earth diameter, driven almost entirely by the count-based \(e\)-like channel. This channel relies on the dominant survival probability \(T_{ee} \simeq 0.99\) and the well-measured antineutrino-to-neutrino cross-section ratio (\(\sigma_{\bar\nu}/\sigma_\nu \simeq 0.49\)). Because these quantities are insensitive to the exact matter density profile, the SFK rate changes by less than 10\% when the three-layer Earth model is replaced by the full tabulated PREM. 

In contrast, the oscillation-phase keying (OPK) scheme relies on the appearance probability \(T_{e\mu}\). At 25 GeV, this is a small interference term heavily suppressed by the MSW effect. While OPK offers the operational advantage of constant-intensity operation, its rate (\(5.2 \times 10^{-6}\) bit/s) is an order of magnitude lower than SFK and is highly fragile. A full PREM evaluation lowers the OPK rate by a factor of three, and a mere \(\pm 2\%\) rescaling of the Earth's density changes it by an order of magnitude. Any practical OPK link would therefore require a near-perfect regional density profile and a narrow-band beam to control energy smearing.

\emph{The \(\nu_\tau\) bottleneck.} The numerical rates are ultimately limited not just by the source flux, but by the available detector technology. At the first oscillation maximum, the beam is almost entirely composed of \(\nu_\tau\) (\(T_{\mu\tau} \simeq 0.98\)). Because current detectors (such as LArTPCs) cannot efficiently identify \(\nu_\tau\) charged-current events, this dominant channel is lost. As shown in Fig.~\ref{fig:tau_bottleneck}, if a future detector could tag \(\nu_\tau\) with 100\% efficiency, the muon-channel contribution to the SFK rate would increase by a factor of five. 

\emph{Comparison with classical intensity keying.} The 2012 NuMI/MINERvA demonstration achieved an OOK rate of \(\sim 0.1\) bit/s over a baseline of \(\sim 1\) km \cite{ref1}. The through-Earth link analyzed here spans \(12,742\) km. Since the event rate for a fixed energy scales as \(R \propto L^{-2}\), the baseline alone reduces the rate by a factor of \(\sim 10^8\). Thus, the low rates we obtain (\(\sim 10^{-4}\) bit/s) are not a failure of flavor modulation, but simply the consequence of the immense distance and the fundamental flux limit. 

\emph{Implications for quantum communication.} It is worth reiterating that the channel analyzed here is strictly classical. While neutrino oscillation is a coherent quantum process, the detection of a specific flavor and charge is a classical measurement. The information is carried by the Poisson statistics of the detected events, not by quantum entanglement or squeezing. Consequently, flavor-modulated neutrino communication offers no quantum-capacity advantage and does not violate the no-signaling theorem. Its value lies entirely in the physical properties of the channel: directional access, constant-intensity operation, and independence from the flux normalization.

The framework presented here provides a rigorous mathematical bridge between three-flavor oscillation physics and classical information theory. It confirms that while neutrinos are impractical for high-throughput telecommunications, they offer a unique and rich signaling paradigm for ultra-low-rate, secure, and robust links.

\section{Practical Considerations}\label{s10}

\emph{Backgrounds.} Atmospheric neutrinos \cite{ref20}, cosmic muons, and intrinsic beam contamination produce \(b_{k}\) in Eq.~\eqref{eq:10}. Gating to the bunch time window of width \(\Delta t\) reduces the atmospheric background by the duty factor \(\Delta t/T_{\rm rep}\). Energy resolution reduces contrast by \(e^{- 2\sigma_{\Phi}^{2}}\) (Sec.~\ref{s6.2}). Density uncertainty \(\delta\rho\) shifts \(A\) and hence \(T_{e\mu}\); to first order \(\delta P_{0} = \partial_{A}P_{0}\delta A\) with \(\delta A/A=\delta\rho/\rho\). As shown in Sec.~\ref{s7.3}, appearance-based schemes (OPK, MPK) are sensitive to the profile, including with the full tabulated PREM and under a \(\pm 2\text{\%}\) density rescaling, while SFK is robust. 

\emph{Baselines.} For \(L/E\) large or \(\sigma_{E}\) wide the oscillations average to \(\langle P_{\alpha\beta}\rangle=\sum_i|U_{\alpha i}|^2|U_{\beta i}|^2\) and flavor modulation offers no advantage. 

\emph{Security and operations.} The beam is directional (\(\theta_{b} \sim 3\) mrad, a footprint fraction \(\sim 10^{- 5}\) of the Earth's surface), so interception requires a detector of comparable mass inside the beam, and jamming requires a symbol-independent neutrino rate exceeding the signal rate at the receiver (Sec.~\ref{s5.8}). Constant-intensity operation makes the modulation invisible to a counting-only observer and removes the dependence on the flux normalization. No cryptographic claim is made.

\section{Conclusion}\label{s11}

The oscillation operator in a layered Earth defines a Poisson-output flavor channel whose performance is governed by the error exponent \(\kappa\) in events per second. At \(E \simeq 25\) GeV on the Earth diameter, matter and the \(\nu_{\mu} \rightarrow \nu_{\tau}\) maximum reduce the informative event fraction. The mean charged-current rate is \(1.31 \times 10^{-2}\,\mathrm{s^{-1}}\), but the information rate derived from the error exponents is \(1.36 \times 10^{-4}\) bit/s for SFK and \(5.2 \times 10^{-6}\) bit/s for OPK with a neutrino-factory-class source.

The lasting value of the framework lies in the physical properties of flavor keying, namely constant-intensity operation, independence of the decision rule from the flux normalization, and directional access, and not in throughput. In summary, the framework provides a mathematical bridge between three-flavor oscillation physics and the theory of Poisson channels. It shows that neutrinos are impractical carriers for high-throughput telecommunication, because the rate is set by the product of source intensity and detector mass, while flavor modulation offers a distinct signaling paradigm: constant-intensity operation that is invisible to a counting-only observer, flux-independent decision making, and directional access. None of these is a cryptographic guarantee. 

We further find that the OPK estimate is sensitive to the Earth density profile: a full tabulated-PREM evaluation lowers it by a factor of about two-and-a-half relative to a three-layer model, and a \(\pm 2\%\) rescaling of the density changes it by an order of magnitude. In contrast, the SFK estimate remains robust, varying by less than 10\% across the same density variations. 

\begin{figure}[tbp]
\centering
\includegraphics[width=0.75\textwidth]{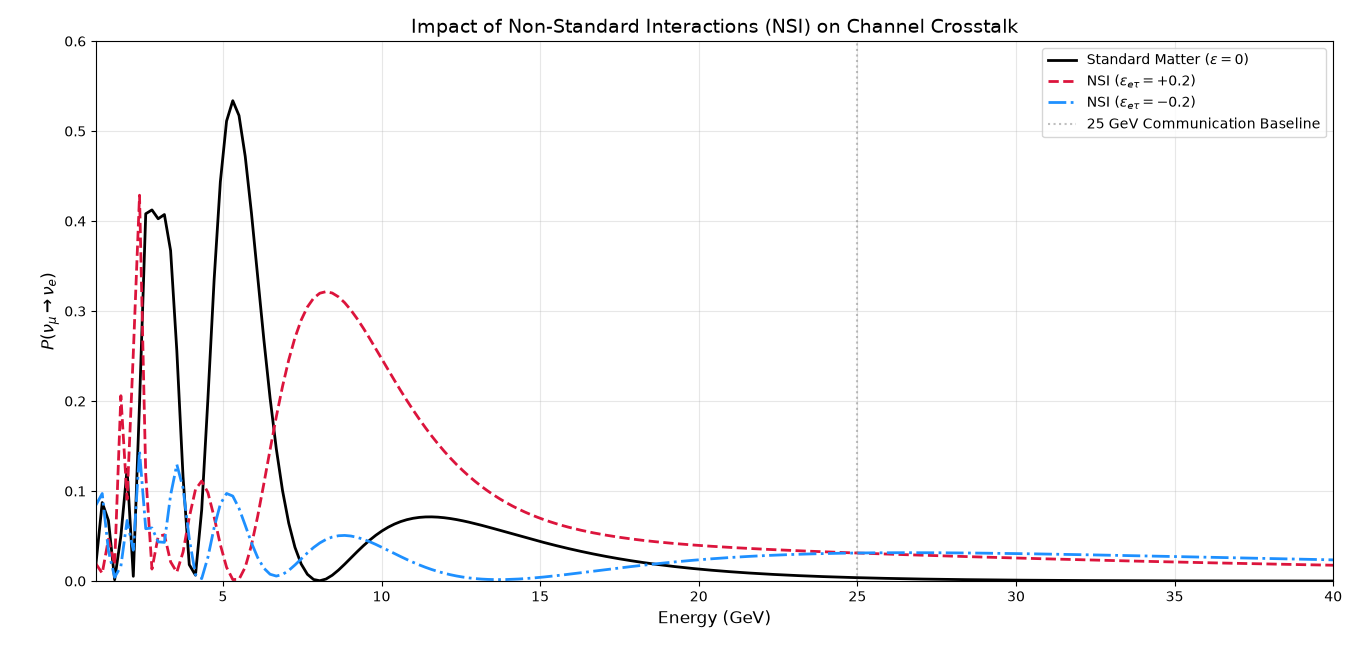}
\caption{Impact of Non-Standard Interactions (NSI) on the channel crosstalk. The probability \(P(\nu_\mu \to \nu_e)\) is shown for the Standard Model (\(\varepsilon = 0\)) and for NSI parameters \(\varepsilon_{e\tau} = \pm 0.2\) as a function of beam energy. The 25 GeV communication baseline is indicated by the dotted vertical line.}
\label{fig:nsi}
\end{figure}

The most promising near-term directions are proof-of-principle SFK links at fixed energy over shorter baselines (where \(R \propto L^{- 2}\)), OPK at the 3--8 GeV matter-resonance region with a narrow-band beam, and flavor as a robustness and security feature for ultra-low-rate links. Open problems are optimal codes for the sparse Poisson-flavor channel, a region-by-region propagation of the PREM density uncertainty (the full tabulated PREM evaluation of Sec.~\ref{s7.3} shows that appearance-based schemes need it, as do the Earth-density effects of Ref.~\cite{ref16}), and the effect of sterile neutrinos or non-standard interactions on \(T_{\alpha\beta}\), including anomalous matter potentials in the inner core \cite{ref17}. The sensitivity of the channel crosstalk to non-standard interactions (NSI) is explicitly demonstrated in Fig.~\ref{fig:nsi}.

\section*{Conflict of Interest}
The authors declare that they have no conflict of interest.

\section*{Funding}
This research received no external funding.

\section*{Ethics Statement}
This study involves numerical simulations and theoretical analysis only. No human participants or animals were involved; therefore, ethical approval was not required.

\section*{Data Availability}
The numerical codes, simulation scripts, and datasets generated during and/or analyzed during the current study are openly available in the Zenodo repository at \url{https://doi.org/10.5281/zenodo.24392721}.

\appendix
\section{Numerical Methodology and Parameters}\label{app:methodology}

This appendix documents the numerical methods and parameter choices used to produce the results in Sections~\ref{s7} and~\ref{s8}. All calculations were performed using double-precision complex linear algebra routines. The baseline parameters are summarized in Table~\ref{tab:pmns} and Table~\ref{tab:earth}, and the numerical checks described below ensure that the analytical approximations used in the main text are robust.

\subsection{PMNS Matrix and Hamiltonian Construction}\label{app:pmns}

The three-flavor evolution is governed by the Hamiltonian in Eq.~\eqref{eq:4}. The PMNS matrix \(U\) is constructed using the standard parameterization:
\begin{equation}
U = \begin{pmatrix}
c_{12}c_{13} & s_{12}c_{13} & s_{13}e^{-i\delta_{CP}} \\
-s_{12}c_{23} - c_{12}s_{13}s_{23}e^{i\delta_{CP}} & c_{12}c_{23} - s_{12}s_{13}s_{23}e^{i\delta_{CP}} & c_{13}s_{23} \\
s_{12}s_{23} - c_{12}s_{13}c_{23}e^{i\delta_{CP}} & -c_{12}s_{23} - s_{12}s_{13}c_{23}e^{i\delta_{CP}} & c_{13}c_{23}
\end{pmatrix},
\end{equation}
where \(c_{ij} = \cos\theta_{ij}\) and \(s_{ij} = \sin\theta_{ij}\). The vacuum Hamiltonian is \(H_{\rm vac} = U \cdot \mathrm{diag}(0, \Delta m^2_{21}, \Delta m^2_{31}) \cdot U^\dagger / (2E)\). The mixing parameters are fixed to the NuFIT 5.3 global fit values for normal ordering, as listed in Table~\ref{tab:pmns}.

\begin{table}[h]
\centering
\begin{tabular}{lc}
\hline
Parameter & Value \\
\hline
\(\theta_{12}\) & \(33.41^\circ\) \\
\(\theta_{13}\) & \(8.58^\circ\) \\
\(\theta_{23}\) & \(49.00^\circ\) \\
\(\delta_{CP}\) & \(232.0^\circ\) \\
\(\Delta m^2_{21}\) & \(7.41 \times 10^{-5}\,\mathrm{eV^2}\) \\
\(\Delta m^2_{31}\) & \(2.507 \times 10^{-3}\,\mathrm{eV^2}\) \\
\hline
\end{tabular}
\caption{Neutrino oscillation parameters used in the numerical simulation (NuFIT 5.3, Normal Ordering).}
\label{tab:pmns}
\end{table}

\subsection{Earth Density Models}\label{app:earth}

Two density profiles are used in this work. For the three-layer model (Sec.~\ref{s7.2}), the baseline is divided into the inner core, outer core, and mantle, with the parameters listed in Table~\ref{tab:earth}. For the full tabulated PREM (Sec.~\ref{s7.3}), the Earth diameter is discretized into \(N=4000\) equal-length slabs. The density at each slab is interpolated from the PREM radial polynomial \cite{ref18}, with the ocean layer replaced by crustal rock (\(\rho = 2.6\,\mathrm{g\,cm^{-3}}\)). The electron fraction is \(Y_e = 0.468\) in the core and \(Y_e = 0.494\) in the mantle, and the matter potential is \(V(x) = \sqrt{2} G_F Y_e(x) \rho(x) / m_N\).

\begin{table}[h]
\centering
\begin{tabular}{lccc}
\hline
Layer & Radius range (km) & \(\rho\) (\(\mathrm{g\,cm^{-3}}\)) & \(Y_e\) \\
\hline
Inner core & \(0 - 1220\) & \(13.0\) & \(0.468\) \\
Outer core & \(1220 - 3480\) & \(11.0\) & \(0.468\) \\
Mantle & \(3480 - 6371\) & \(4.5\) & \(0.494\) \\
\hline
\end{tabular}
\caption{Three-layer Earth model parameters used in the numerical calculation of Sec.~\ref{s7.2}.}
\label{tab:earth}
\end{table}

\subsection{Numerical Evolution and Transition Matrices}\label{app:evolution}

For a given energy \(E\) and path \(P\), the total evolution operator is \(S(E;P) = \mathcal{T}\exp[-i\int_P H(x)\,dx]\). In the three-layer model, this reduces to the product of three constant-density evolution operators:
\begin{equation}
S_{\rm 3-layer} = \exp(-i H_{\rm mantle} L_{\rm mantle}) \exp(-i H_{\rm outer} L_{\rm outer}) \exp(-i H_{\rm inner} L_{\rm inner}),
\end{equation}
where each \(H_k = H_{\rm vac} + \mathrm{diag}(V_k, 0, 0)\). In the PREM model, the evolution operator is computed as a product over \(N\) slabs:
\begin{equation}
S_{\rm PREM} = \prod_{j=N}^{1} \exp\left(-i H(x_j) \Delta x\right),
\end{equation}
where \(\Delta x = D_{\oplus}/N\). The transition probabilities are then \(T_{\alpha\beta} = |(S)_{\beta\alpha}|^2\). The numerical stability of the PREM calculation was verified by varying \(N\) between 1000 and 4000; the transition probabilities at 25 GeV change by less than 1\%, as quoted in Sec.~\ref{s7.3}.

\subsection{Validation of the Bhattacharyya Bound}\label{app:chernoff}

In Sec.~\ref{s4.1}, we use the closed-form Bhattacharyya exponent \(E_B\) (Eq.~\ref{eq:13}) to compute the information rates, and we claim that this bound is tight. To validate this claim, we numerically optimized the exact Chernoff exponent \(E_C\) (Eq.~\ref{eq:14}) for the SFK \(e\)-like channel at the nominal symbol time. For expected counts \(\lambda_0 = 36.4\) and \(\lambda_1 = 74.9\), the optimization yields:
\begin{align}
E_C &= 3.4478, \\
E_B &= 3.4354, \\
\frac{E_C}{E_B} &= 1.0036.
\end{align}
The \(0.4\%\) difference confirms that the analytical bound is an accurate proxy for the exact maximum-likelihood limit, and that the bit rates reported in Sections~\ref{s7.4} and~\ref{s8.2} are not merely conservative lower bounds.

\subsection{Cross-Section and Detector Assumptions}\label{app:detector}

The per-species CC rate in the absence of oscillation is \(R = \Phi' \sigma(E) N_t\), with \(\sigma(E) = 0.68 \times 10^{-38} E\,\mathrm{cm^2}\) per nucleon \cite{ref19}. For the Earth-diameter link (Sec.~\ref{s7.2}), the detector mass is \(N_t = 1.2 \times 10^{35}\) nucleons (200 kt), and the source intensity is \(N_{dec} = 3 \times 10^{13}\,\mathrm{s^{-1}}\) \cite{ref8}. The antineutrino-to-neutrino cross-section ratio is fixed to \(\sigma_{\bar\nu}/\sigma_\nu = 0.49\). The detector confusion matrix \(\Gamma\) is modeled as diagonal with efficiency \(\varepsilon = 0.9\), and the charge-misidentification probability is \(\delta_c = 10^{-3}\). The target error probability for all rate calculations is \(P_e = 10^{-2}\).

\end{document}